\documentclass[10pt]{iopart}
\usepackage{theorem}
\usepackage{iopams}
\usepackage{graphicx}
\usepackage{dsfont}
\usepackage{bm}
\usepackage{mathrsfs}
\usepackage[T1]{fontenc}
\usepackage[usenames]{color}
\usepackage{hyperref}

\newcommand{\lvert}{|}
\newcommand{\rvert}{|}
\newcommand{\abs}[1]{\left\lvert #1\right\rvert}
\newcommand{\avg}[1]{\left\langle #1 \right\rangle}

\newcommand\CC{{\mathds{C}}}
\newcommand\one{{\mathds{1}}}
\newcommand\eps{{\varepsilon}}

\newcommand{\Lapt}{{\mathscr L}}
\newcommand{\Lu}{\hat{u}}

\newcommand{\setP}{\mathfrak{P}}

\newcommand{\seta}{\mathfrak{a}}
\newcommand{\cfra}[2]{\frac{#1}{#2+}}
\newcommand{\tmat}{T} 
\newcommand{\trsp}{\tmat} 
\newcommand{\tmb}{m}
\newcommand{\tm}{m}
\newcommand{\prs}{{\bm\sigma}}
\renewcommand{\le}{\leqslant}
\renewcommand{\ge}{\geqslant}
\newcommand{\pb}[1]{\left[#1\right]} 

\begin{document}
\title{Equilibration in long-range quantum spin systems from a BBGKY perspective}
\author{Rytis Pa\v{s}kauskas$^1$ and Michael Kastner$^{1,2}$}
\address{$^1$ National Institute for Theoretical Physics (NITheP), Stellenbosch 7600, South Africa}
\address{$^2$ Institute of Theoretical Physics, University of Stellenbosch, Stellenbosch 7600, South Africa}
\ead{\mailto{rytis.paskauskas@gmail.com},\mailto{kastner@sun.ac.za}}

\begin{abstract}
The time evolution of $\ell$-spin reduced density operators is studied for a class of Heisenberg-type quantum spin models with long-range interactions. In the framework of the quantum Bogoliubov-Born-Green-Kirkwood-Yvon (BBGKY) hierarchy, we introduce an unconventional representation, different from the usual cluster expansion, which casts the hierarchy into the form of a second-order recursion. This structure suggests a scaling of the expansion coefficients and the corresponding time scales in powers of $N^{1/2}$ with the system size $N$, implying a separation of time scales in the large system limit. For special parameter values and initial conditions, we can show analytically that closing the BBGKY hierarchy by neglecting $\ell$-spin correlations does never lead to equilibration, but gives rise to quasi-periodic time evolution with at most $\ell/2$ independent frequencies. Moreover, for the same special parameter values and in the large-$N$ limit, we solve the complete recursion relation (the full BBGKY hierarchy), observing a superexponential decay to equilibrium in rescaled time $\tau=tN^{-1/2}$.
\end{abstract}

\date{\today}

\section{Introduction}\label{s:intro}
Equilibration and thermalization in closed quantum systems have recently seen renewed interest, triggered by the impressive progress in performing experiments with ultracold atoms and ions \cite{BlochDalibardZwerger08}. In these experiments the coupling to the environment is negligible on the accessible timescales, and they can therefore be regarded to a very good approximation as closed systems. Typically, equilibration (in a suitable sense) is expected to take place in closed quantum systems \cite{vonNeumann29,Reimann08,Goldstein_etal10}, but important exceptions are known to exist. Failure of equilibration in systems close to integrability has been observed experimentally by Kinoshita {\em et al.} \cite{Kinoshita_etal06}, and a substantial body of theoretical and numerical work has been devoted to this problem over the last years (see \cite{Polkovnikov_etal11} for a review).

Studying the long-time dynamics of many-body quantum systems in general is a daunting task. For numerical work, one is usually restricted to fairly small system sizes, much smaller than most experimental realizations and not anywhere close to the macroscopic regime. The situation is different for integrable systems where analytic solutions may exist and larger systems may be studied. However, these systems are known to show atypical behavior, possibly very different from the non-integrable systems one is often interested in \cite{Rigol_etal07}.

One solvable model (a quantum spin model with Ising-type interactions and subjected to an external magnetic field) for which the long-time evolution can be studied analytically, was proposed by Emch \cite{Emch66} in 1966 and later studied by Radin \cite{Radin70} in more detail. 
For a certain class of initial conditions, the time evolution of the expectation values of operators $A$ of a certain kind can be calculated analytically for arbitrary system sizes $N$, showing non-Markovian relaxation to the thermal average in the thermodynamic limit $N\to\infty$ \cite{Emch66}. This model has been extended by one of us to the case of long-range interactions, i.e.\ spin-spin interactions decaying like $r^{-\alpha}$ with the distance $r$ on a $d$-dimensional lattice, where the exponent $\alpha$ satisfies $0\le\alpha\le d$ \cite{Kastner11,KastnerCEJP}. As in the short-range case, the expectation values of the operators $A$ relax to thermal equilibrium, but they do so on a time scale that diverges with the system size $N$: The larger the system is, the longer it takes to thermalize. In fact, the relaxation dynamics becomes so slow that, within a given experimental resolution and for large enough $N$, no deviation from the initial state will be observed.

A similar phenomenon, going under the name of quasi-stationary states, has received considerable attention in the field of classical long-range interacting systems \cite{CamDauxRuf09,BouchetGuptaMukamel10}, and in particular in classical gravity \cite{JoyceWorrakitpoonpon10}. In the classical context, kinetic theory, and in particular the Vlasov (or the collisionless Boltzmann) equation, has been successfully applied to describe these quasi-stationary states \cite{Barre_etal06,Antoniazzi_etal07,CamDauxRuf09,BouchetGuptaMukamel10,ChavanisBaldovinOrlandini11}. Such a Vlasov description is known to become exact for long-range systems in the thermodynamic limit, and is therefore also expected to be a good starting point for a description of large but finite systems \cite{BraunHepp77}. The main goal of the present work is to study, similarly to the classical case, the long-time persistence of quasi-stationary non-equilibrium states and their final relaxation to equilibrium for long-range quantum spin systems within the framework of quantum kinetic theory. We can then test the predictions of the quantum kinetic theory against the analytic result for the Emch-Radin model reported in \cite{Kastner11,KastnerCEJP}, but the scope of such a kinetic theory goes beyond this model: On the basis of a tested and well-working kinetic theory, equilibration can then be studied for non-integrable generalizations of the Emch-Radin model, or non-integrable long-range variants of the Heisenberg model for which exact solutions are not known.

In addition to providing a tool for investigating the dynamics of quantum spin models, our results also shed light on more general features regarding the role of closure conditions when truncating the BBGKY hierarchy: The particularly simple structure of spin-$1/2$ lattice models facilitates analytic calculations beyond what can be achieved in continuum systems, and an understanding of the effect of approximation schemes is easier to attain.

The article is structured as follows: In \sref{s:spinmodel} we introduce a fairly general class of long-range interacting quantum Heisenberg models with anisotropic interactions, subjected to an external magnetic field, and discuss its relation to the exactly solvable long-range Emch-Radin model. A short introduction to the quantum BBGKY hierarchy is given in \sref{s:BBGKY}, with special emphasis paid to the role of closure conditions when truncating the hierarchy. In \sref{s:definition}, we introduce a representation of the hierarchy, i.e.\ a certain choice of basis in the underlying Hilbert space, different from the conventional cluster expansion. In \sref{s:BBGKYexpansion}, the BBGKY hierarchy is expressed in terms of this representation, and we find that it has the form of a second-order recursion relation. This structure turns out to be crucial for the derivation of our results. First, as discussed in \sref{s:thermodynamic}, the recursion relation suggests a scaling of the expansion coefficients and the corresponding time scales in powers of $N^{1/2}$ with the system size $N$, implying a separation of time scales in the large system limit. The largest time scale is found to diverge proportionally to $N^{1/2}$ in the large system limit. As a consequence, equilibration is expected on a time scale that diverges in the thermodynamic limit, in agreement with what is known about the long-range Emch-Radin model. In \sref{s:case}, we consider the long-range anisotropic quantum Heisenberg model for a special set of parameter values and initial conditions for which calculations are easier to perform. Under these restrictions, we can show analytically in \sref{s:specialfinite} that closing the BBGKY hierarchy by neglecting $\ell$-spin correlations does never lead to equilibration, but gives rise to quasi-periodic time evolution with at most $\ell/2$ independent frequencies. Moreover, in \sref{s:speciallimit} we solve the untruncated recursion relation (full BBGKY hierarchy) in the large-$N$ limit, observing a superexponential decay to equilibrium in rescaled time $\tau=tN^{-1/2}$. This behavior and the observed time scale are exact analytic results which are in a perfect agreement with the Emch-Radin model. A more detailed discussion of these results and their implications on the general considerations of \sref{s:thermodynamic} is given in \sref{s:discussion}, and we summarize our findings in \sref{s:conclusions}.

\section{Long-range anisotropic quantum spin model}
\label{s:spinmodel}
Two quantum spin models are introduced in this section. The first one, an anisotropic quantum Heisenberg model with Curie-Weiss type interactions, is the one actually studied in this work. The second one, named after Emch and Radin, is an Ising type quantum spin model in a longitudinal magnetic field, with two-body interactions that decay with the distance as a power law. For special choices of the parameters, both models agree, and the known exact results on the dynamics of the Emch-Radin model can be used to test the validity of our results regarding relaxation to equilibrium in the anisotropic quantum Heisenberg model.

\subsection{Curie-Weiss anisotropic quantum Heisenberg model}
Consider $N$ identical spin-$1/2$ particles, attached to the sites of a $d$-dimensional lattice. The corresponding quantum dynamics takes place on the Hilbert space
\begin{equation}\label{e:Hilbertspace}
  \mathscr{H}_N=\bigotimes_{i=1}^N \CC_i^2,
\end{equation} 
where the $\CC_i^2$ are identical replicas of the two-dimensional Hilbert space of a single spin-$1/2$ particle. The unitary time evolution on $\mathscr{H}_N$ is generated by the $N$-body Hamiltonian
\begin{equation}\label{e:Hamiltonian}
  H_{1\dots N}=\sum_{i=1}^N H_i + \sum_{\scriptstyle i,j=1\atop \scriptstyle i<j}^N V_{ij}
\end{equation}
consisting of an on-site potential and a spin-spin interaction potential,
\begin{equation}\label{e:V12}
  H_i=-\sum_{a\in\mathcal{I}} h^a\sigma_i^a,\qquad V_{ij}=-\frac{1}{N}\sum_{a,b\in\mathcal{I}} J^{ab}\sigma_i^a \sigma_{j}^b,
\end{equation}
where $\sigma_i^a$ denotes the Pauli operators at a lattice site $i$ and
\begin{equation}
\mathcal{I}=\{x,y,z\}
\end{equation}
is the set of component indices.
The spin-spin interaction $V_{ij}$ is of Curie-Weiss type, coupling each spin to every other on the lattice, and the strength of the coupling is determined by a $3\times3$ matrix $J$ of coupling constants with matrix elements $J^{ab}$. The $1/N$ prefactor in \eref{e:V12} is introduced to render the energy per spin finite in the thermodynamic limit. The long-range interactions, and the $N$-dependent prefactor they are necessitating, can lead to peculiar properties. In equilibrium, it may happen that different statistical ensembles (like the microcanonical and the canonical one) are nonequivalent in the thermodynamic limit, and issues of this kind have been discussed in \cite{Kastner10,KastnerJSTAT10}. In the present article, we will focus on peculiarities of the non-equilibrium behavior of this model. 

\subsection{Emch-Radin model}
The Emch-Radin model \cite{Emch66,Radin70} has a Hamiltonian similar to \eref{e:V12}, but with a two-body potential
\begin{equation}
V_{ij}=-\frac{1}{N}J_{ij}^{zz}\sigma_i^z \sigma_{j}^z
\end{equation}
So in contrast to the anisotropic Heisenberg model's two-body potential \eref{e:V12}, $J^{zz}$ is the only nonzero element of the coupling matrix $J$. Moreover, the spins are associated to the sites of a $d$-dimensional lattice, and the coupling between spins $\sigma_i^z$ and $\sigma_{j}^z$ depends algebraically on their distance $D(i,j)$ on the lattice, i.e.\ $J_{ij}^{zz}\propto D(i,j)^{-\alpha}$ with some nonnegative exponent $\alpha$. The on-site potential potential is of the same form as in \eref{e:V12}, but with an external magnetic field $h=(0,0,h^z)$ pointing in $z$-direction. It follows from these definitions that the Hamilton operators of the anisotropic Heisenberg model and the Emch-Radin model have a special case in common: For $\alpha=0$ the distance dependence of the Emch-Radin coupling $J_{ij}^{zz}$ is eliminated and the Hamiltonian is identical to that of the Curie-Weiss anisotropic quantum Heisenberg model with $J=\mbox{diag}(0,0,J^{zz})$ and $h=(0,0,h^z)$.

\begin{figure}\center
\includegraphics[width=0.55\linewidth]{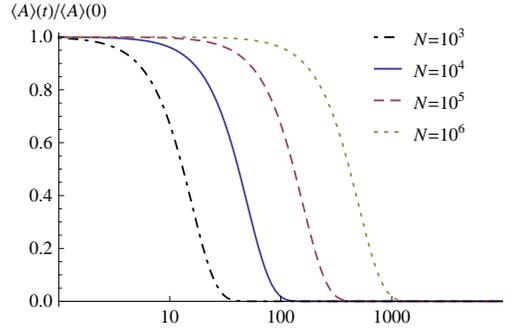}
\caption{\label{f:decay}
Time evolution of the expectation value of the observable $A$ for magnetic field $h=0$, coupling matrix $J=\mbox{diag}(0,0,1)$, and exponent $\alpha=0$. The plot was obtained by evaluating with {\sc Mathematica} the analytic formula (9) of reference \cite{Kastner11} for various system sizes $N$. The result is valid for arbitrary initial states and arbitrary coefficients $a_i$ in the expansion \eref{eq:A} of $A$. The expectation value shows an apparent decay towards its zero equilibrium value, but on a time scale that depends strongly on the system size $N$ (note the logarithmic time scale). The decay is only apparent as, on time scales much longer than shown, recurrent behavior (Loschmidt echos) occur due to the system's finite size. Similar behavior is observed for other values of $\alpha$ between zero and the lattice dimension $d$.
}
\end{figure}

The permissible initial states in the Emch-Radin model are restricted to density operators which are diagonal in the $\sigma^x$ tensor product eigenbasis of $\mathscr{H}_N$. Under this condition, an analytic expression can be obtained for the time evolution of expectation values of operators of the form
\begin{equation}\label{eq:A}
A=\sum_{i=1}^N a_i \sigma_i^x
\end{equation}
with real coefficients $a_i$ \cite{Emch66}. In the case of short-range interactions, i.e.\ for exponents $\alpha>d$, the time evolution of the expectation values $\langle A\rangle(t)$ was analyzed in the thermodynamic limit $N\to\infty$ in \cite{Emch66}. The system was found to relax to equilibrium, and the process of relaxation was shown to be non-Markovian. The long-range scenario with $0\le\alpha\le d$, analyzed in \cite{Kastner11,KastnerCEJP}, was shown to display a remarkably different behavior: The time scale at which equilibration appears to occur was found to increase proportionally to $N^r$ with $r=\min\{1/2,1-\alpha\}$. This finding implies a diverging equilibration time scale in the thermodynamic limit (see \fref{f:decay} for an illustration). 

The analytic results in \cite{Emch66} and \cite{Kastner11,KastnerCEJP} are not easily extended to other parameter values, different initial conditions, or more general observables. We believe, however, that similar quasi-stationary behavior (i.e.\ long-lived non-equilibrium behavior per\-sist\-ing on a time scale that diverges with increasing system size) shows up under much more general conditions. The aim of this work is to employ quantum kinetic theory to study the long-time dynamics of long-range interacting quantum spin systems of the Heisenberg type \eref{e:Hamiltonian} and \eref{e:V12}. For the moment, we restrict the analysis to Curie-Weiss-type potentials as in \eref{e:V12}, but we expect qualitatively similar results to hold for algebraically decaying long-range potentials with exponents $\alpha$ between zero and the lattice dimension $d$.

\section{BBGKY hierarchy}
\label{s:BBGKY}
The density operator $\rho_N$ of an $N$-spin system is a self-adjoint, positive, trace-class operator, acting on the Hilbert space $\mathscr{H}_N$. We use the normalization convention 
\begin{equation}\label{e:DOP}
  \Tr_{1\ldots N} \rho_N = 1,
\end{equation}
where $\Tr_{1\ldots N}$ denotes a trace over all $N$ factors of the tensor product Hilbert space \eref{e:Hilbertspace}. The expectation value of an operator $A$ with respect to $\rho_N$ is given by
\begin{equation}
\langle A\rangle=\Tr_{1\ldots N}\left(\rho_NA\right).
\end{equation} 
The time evolution of the density operator is governed by the von Neumann equation
\begin{equation}\label{e:vn}
  \rmi\hbar\partial_t \rho_N=\pb{H_{1\ldots N},\rho_N}.
\end{equation}
Reduced $\ell$-particle density operators are derived from $\rho_N$ by tracing out $(N-\ell)$ of the factors of $\mathscr{H}_N$,
\begin{equation}\label{e:F}
  F_{1\dots\ell}=\Tr_{\ell+1\dots N}\rho_N,
\end{equation}
where $\Tr_{\ell+1\dots N}$ denotes such a partial trace. The $F_{1\ldots\ell}$ are again density operators, i.e.\ self-adjoint, positive operators of trace-class. Since they are all derived from the same $\rho_N$, the reduced density operators are not independent, but satisfy a collection of consistency conditions of the form
\begin{equation}\label{e:traceprop}
\Tr_\ell F_{1\dots\ell}=F_{1\dots\ell-1}
\end{equation}
(with the convention of $F_0\equiv1$) which we will refer to as the {\em trace property} of $F_{1\ldots\ell}$. 

We will assume in the following that all reduced density operators $F_{1\ldots\ell}$ are invariant under $\ell$-permutations of their indices from the set $\{1,\dots,N\}$. For $\ell=1$, for example, this means simply that
\begin{equation}
F_1=F_2=\cdots=F_N,
\end{equation}
and for $\ell=2$ this property amounts to
\begin{equation}
F_{ij}=F_{kl}\qquad\forall i,j,k,l\in\{1,\dots,N\},\quad i\neq j\,,\quad k\neq l.
\end{equation}
In the context of continuum (off-lattice) systems, permutation invariance is usually justified by the indistinguishability of particles. For the lattice systems investigated in the present article, the spin degrees of freedom can of course be distinguished by the lattice site they are attached to. Instead, in that context the assumption of permutation invariance amounts to assuming that the system is in a homogeneous state, i.e.\ different subsystems behave similarly. This assumption is expected to be justified in the Heisenberg model \eref{e:V12} with ferromagnetic coupling $J^{ab}\ge0$, but will be violated in the presence of anti-ferromagnetism. Since the Hamiltonian \eref{e:Hamiltonian} and \eref{e:V12} is permutation invariant as well, the homogeneity of an initial state will be preserved under time evolution. 

Under the assumption of permutation invariance, the Bogoliubov-Born-Green-Kirkwood-Yvon (BBGKY) hierarchy is obtained by applying the partial trace $\Tr_{\ell+1\ldots N}$ to the von Neumann equation \eref{e:vn}. For a Hamiltonian of the form \eref{e:Hamiltonian}, the hierarchy can be written as
\begin{equation}\label{e:spinkin}
  \fl \rmi\hbar\partial_t F_{1\ldots\ell} =
  \sum_{i=1}^{\ell}\pb{H_i,F_{1\ldots\ell}} + \sum_{\scriptstyle i,j=1\atop \scriptstyle i<j}^\ell\pb{V_{ij},F_{1\ldots\ell}}+ (N-\ell) \Tr_{\ell+1}\sum_{i=1}^{\ell}\pb{V_{i,\ell+1},F_{1\ldots\ell+1}}.
\end{equation}
For a given number $N$ of spins, this is a finite set of equations, and it is fully equivalent to the von Neumann equation \eref{e:vn} from which it was derived: A solution of the BBGKY hierarchy is equivalent to an exact solution for the density operator $\rho_N$, from which the reduced density operators $F_{1\ldots\ell}$ can be obtained via equation \eref{e:F}.

\subsection{Closure conditions}
\label{s:closure}
In the form \eref{e:spinkin}, the hierarchy is not yet particularly useful: Solving the full hierarchy is as difficult as solving the von Neumann equation (and therefore impossible for most cases of interest). For many physical problems, it turns out that $n$-particle correlations are naturally ordered by decreasing relevance, and only those with $n\le\ell$ have to be considered (where $\ell$ is some small number, usually not larger than 4, depending on the system under investigation and the quantity of interest). It is therefore often sufficient to consider only the time evolution of the first $\ell$ reduced density operators. Hence a useful computational tool may be derived by truncating the hierarchy at the level of the $\ell$-spin reduced density operator. However, the achieved simplification of such a truncation comes at the expense of an approximation. 

Simply truncating the set \eref{e:spinkin} after the first $\ell$ equations results in an ill-defined problem: The $\ell$th equation, which determines the time evolution of $F_{1\ldots\ell}$, requires $F_{1\ldots\ell+1}$ as an input, but the equation determining $F_{1\ldots\ell+1}$ has been eliminated by truncation. To obtain a well-defined system of $\ell$ equations, a {\em closure condition}\/ has to be postulated between $F_{1\ldots\ell+1}$ and the lower order reduced density operators. We refer to such a relation among $F_{1\ldots\ell+1}$ and all of the $\{F_{1\ldots j}\}_{j=1}^{\ell}$ as an $\ell$th order closure condition. 

A frequently used truncation scheme of the BBGKY hierarchy is based on the so-called cluster expansion \cite{Bonitz}, where correlation operators $G_{1\ldots\ell}$ are defined according to the following scheme,
\numparts
\begin{eqnarray}
  F_{12}&=&F_1F_2+G_{12} ,\label{e:G2}\\
  F_{123}&=&F_1F_2F_3+F_1G_{23}+F_2G_{31}+F_3G_{12}+G_{123}\label{e:G3},\\
  F_{1234}&=&F_1F_2F_3F_4+F_1F_2G_{34}+F_1F_3G_{24}+F_1F_4G_{23}+F_2F_3G_{14}\nonumber\\
  &&+F_2F_4G_{13}+ F_3F_4G_{12}+G_{12}G_{34}+G_{13}G_{24}+G_{14}G_{23}\nonumber\\
  &&+F_1G_{234}+F_2G_{134}+F_3G_{124}+F_4G_{123}+G_{1234},\label{e:G4}
\end{eqnarray}
\endnumparts
and so on. Based on this expansion, a straightforward way to close the BBGKY hierarchy at $\ell$th order is to express \eref{e:spinkin} in terms of the correlation operators and set
\begin{equation}\label{e:correlationclosure}
G_{1\ldots\ell+k}=0\qquad\forall\, 1\le k\le N-\ell\,.
\end{equation}
We shall refer to this approximation as the $\ell$th order correlation closure.

Needless to say that the accuracy of the approximation  will depend on the ``quality'' of the closure condition. Regardless of the details of the closure, it seems plausible to expect an improvement in the accuracy with increasing order $\ell$ of the truncation. We have tested this expectation by numerically investigating the BBGKY hierarchy truncated by correlation closures of various orders. The density operators were expanded in the basis of Pauli matrices and the evolution was performed for the coefficients of the expansion (discussed in detail in \sref{s:representation}). The one-spin reduced density operator in this expansion reads
\begin{equation}\label{e:F1expansion}
F_1=\frac{1}{2}\biggl(\one_1+\sum_{a\in\mathcal{I}} f_1^a \sigma_1^a\biggr),
\end{equation}
where $\one_1$ denotes the identity operator on the one-spin Hilbert space. The real expansion coefficients $f_1^a$ are related to the mean spin expectation value, as
\begin{equation}
\overline{\sigma^a}\equiv \avg{\frac{1}{N}\sum_{i=1}^N\sigma_i^a}
  =\Tr_{1\ldots N}\frac{1}{N}\sum_{i=1}^N\sigma_i^a\rho_N
  =\Tr_1\sigma_1^aF_1=f_1^a.
\end{equation}
The time evolution of the modulus
\begin{equation}
\abs{f_1}=\sqrt{(f_1^x)^2+(f_1^y)^2+(f_1^z)^2}
\end{equation}
is displayed in \fref{f:diverg} for correlation closures of orders $\ell=2,3,4$. To summarize the simulations, the effect of increasing the order of the truncation is rather disastrous and the long-time evolution is badly predicted at either order:
\begin{figure}
  \center
  \includegraphics[width=0.6\linewidth]{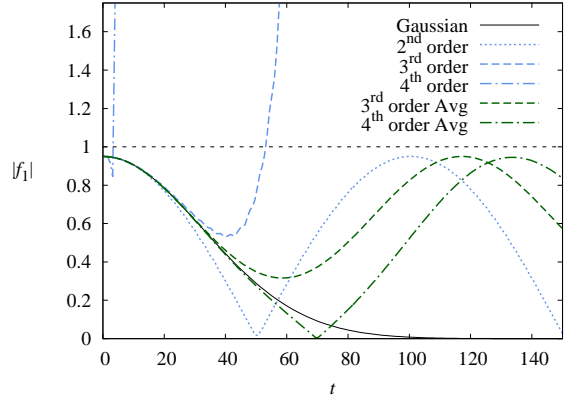}
  \caption{\label{f:diverg}Time evolution of the modulus $|f_1|$ of the coefficient vector $f_1$ parameterizing the one-spin reduced density operator $F_1$. The different curves correspond to correlation closures of orders $\ell=2$, 3, and 4. The parameter values in the Hamiltonian \eref{e:Hamiltonian} are $h=(0,0,1)$, $J=\mathrm{diag}(0,0.04,1)$, and $N=400$, the initial conditions are chosen as $f_1(0)=(0.95,0,0)$ and $g_\ell(0)=0$ for $\ell\ge2$ [see \eref{e:g2}--\eref{e:g4}]. The constant line at 1 indicates the boundary value of condition \eref{e:f1pos} at which $F_1$ ceases to be a positive operator. Remarkably, the larger the order $\ell$ of the correlation closure, the earlier this property is violated. Using, instead of the original equations, {\em averaged}\/ (Avg) time evolution equations as introduced in \sref{s:thermodynamic} where the fast oscillations have been eliminated, the numerics improves significantly in the sense that $F_1$ remains a positive operator for the times shown. The Gaussian plotted as a solid line indicates the behavior expected from the exact time evolution in the $N\to\infty$ limit.}
\end{figure}
According to definition \eref{e:F}, density operators $F_{1\ldots\ell}$ have to be positive operators, and this property is conserved by the von Neumann equation \eref{e:vn}. In the Pauli representation \eref{e:F1expansion}, positivity of $F_1$ amounts to the condition
\begin{equation}\label{e:f1pos}
  \abs{f_1} \le 1
\end{equation}
which, as is evident from \fref{f:diverg}, is violated for the correlation closures of order $\ell\geqslant3$ after relatively short times. So the higher-order correlation closures not only fail to improve on the lower-order ones regarding the relaxation to equilibrium, but they even fail to preserve the basic features of the density operators. We will see in the forthcoming sections that the latter is an artefact of numerics, caused by the presence of oscillatory degrees of freedom with several fundamentally different time scales. This problem will be dealt with in \sref{s:thermodynamic}, where an averaging procedure is defined that eliminates the fast oscillations while correctly reproducing the evolution on the slow time scale.

It is the main objective of the present paper to investigate and better understand the effect of the correlation closures at various orders, and in particular the question of whether and how this type of closure condition can correctly describe the relaxation to equilibrium in a long-range quantum spin system in the thermodynamic limit.

\section{Representation of the {BBGKY} hierarchy}
\label{s:representation}
The derivation of the results reported in this article depends crucially on a particular expansion of the reduced $\ell$-spin density operators in the basis of Pauli operators. Expressed in terms of this expansion, the BBGKY hierarchy reveals a recursive structure which is at the basis of the analytical results obtained.

\subsection{Definition of the expansion}
\label{s:definition}
Inspecting the BBGKY hierarchy \eref{e:spinkin}, one might at a first glance get the impression that the reduced density operators $F_{1\ldots\ell}$ are coupled by a first order recursion, in the sense that the time evolution equation of $F_{1\ldots\ell}$ contains only one further reduced density operator, $F_{1\ldots\ell+1}$. However, this is not really true, as in fact the various $F_{1\ldots\ell}$ are dependent on each other through trace properties \eref{e:traceprop}. The crucial idea behind the expansion introduced in this section is to choose the expansion coefficients such that the trace property is automatically satisfied. Then the resulting coefficients are independent variables, on the basis of which the ``true'' structure of the equations can be studied.

Expanding the first few reduced density operators in the Pauli basis $\{\one,\sigma^x,\sigma^y,\sigma^z\}$, one can verify by inspection that the following choice of expansion coefficients satisfies the trace properties \eref{e:traceprop},
\numparts
\begin{eqnarray}
  F_1 &=& \frac12\Bigl(\one_1 + \sum_{a\in\mathcal{I}}f_1^a\sigma_1^a\Bigr),\label{e:f1}\\
  F_{12} &=& \frac14\Bigl(\one_{12} +
  \sum_{a\in\mathcal{I}} f_1^a(\sigma_1^a+\sigma_2^a)+\sum_{a,b\in\mathcal{I}}f_2^{ab}\sigma_1^a\sigma_2^b\Bigr),\label{e:f2}\\
  F_{123} &=& \frac18\Bigl(
  \one_{123} + \sum_{a\in\mathcal{I}}f_1^a(\sigma_1^a+\sigma_2^a+\sigma_3^a)\nonumber\\
  &&+\sum_{a,b\in\mathcal{I}}f_2^{ab}(\sigma_1^a\sigma_2^b+\sigma_2^a\sigma_3^b+\sigma_3^a\sigma_1^b)+\sum_{a,b,c\in\mathcal{I}}f_3^{abc}\sigma_1^a\sigma_2^b\sigma_3^c\Bigr).\label{e:f3}
\end{eqnarray}
\endnumparts
We introduce several definitions to facilitate the general formulation. For the purpose of counting the particle (subscript) indices, we define, for $1\le n\le\ell\le N$, the set $\setP_n(\ell)$ consisting of all $n$-element permutations $\left(p_1,\ldots,p_n\right)$ of particle labels $p_i\in\{1,\dots,\ell\}$ such that $p_i<p_j$ for all $1\le i<j\le n$ (i.e.\ all sequences are strictly increasing). Considering for example $n=2$, we have $\setP_2(\ell)=\{(1,2),(1,3),\ldots,(1,\ell),(2,3),\ldots,(\ell-1,\ell)\}$. Moreover, we use the notation 
\begin{equation}
  \prs_p^a=\prod_{i=1}^n \sigma_{p_i}^{a_i}
\end{equation}
with $p\in\setP_n(\ell)$ and the component multi-index $a=(a_1,a_2,\ldots,a_n)$ with $a_i\in \mathcal{I}$. With these definitions, the general pattern behind \eref{e:f1}--\eref{e:f3} can be captured by the formula
\begin{equation}\label{e:Fexp}
  F_{1\ldots\ell} = 2^{-\ell}\sum_{n=0}^{\ell}
  \sum_{a\in \mathcal{I}^n} f_n^{a}
  \sum_{p\in \setP_n(\ell)}\prs_{p}^{a},
\end{equation}
with the convention $f^a_0\equiv1$. 

By inspection of the last terms in \eref{e:f2} and \eref{e:f3}, it follows that $f_2^{ab}$ and $f_3^{abc}$ have to be symmetric with respect to permutations of their superscript indices in order to make sure that the resulting expressions for $F_{12}$ and $F_{123}$ are symmetric under particle exchange. As a result, not all components $f_\ell^a$ are independent variables. In particular, any element $f_\ell^{a'}$ is equal to the element $f_\ell^{a(m)}$, where
\begin{equation}\label{e:a(m)}
a(m)=(\underbrace{x,\ldots,x}_{\hspace{-3mm}\displaystyle \tm_x\mbox{ elements}\hspace{-3mm}},y,\ldots,y,z,\ldots,z)
\end{equation}
is a permutation of $a'$ with the components $x$, $y$, $z$ arranged in contiguous blocks of $\tm_x$, $\tm_y$, and $\tm_z$ labels. Hence, all independent components of $f_\ell$ can be labelled uniquely by a triple of nonnegative integers
\begin{equation}
  m=(\tm_x,\tm_y,\tm_z),\quad \tm_i\ge 0,
\end{equation}
where $m_x+m_y+m_z=\ell$. In the following, notation in terms of $a$ and $\tmb=\tmb(a)$ will be considered as equivalent, $f_\ell^a\equiv f_\ell^\tmb$. 

The number of independent components of $f_\ell^\tmb$ is equal to $d(\ell)=(\ell+1)(\ell+2)/2$. Therefore the $N$-spin density operator is described by
\begin{equation}\label{e:dofcount}
  D(N)=\sum_{\ell=1}^{N}d(\ell)=\frac16(N+1)(N+2)(N+3)-1
\end{equation}
independent real variables, a number substantially smaller than the $4^N$ real variables necessary to parameterize an arbitrary Hermitian operator on the $N$-spin Hilbert space $\mathscr{H}_N$ given in \eref{e:Hilbertspace}.

\subsection{Relation to the cluster expansion}
In order to implement the correlation closures \eref{e:correlationclosure}, we need to express the correlation operators $G_{1\ldots\ell}$ in terms of the expansion coefficients $f_\ell^{a_1\ldots a_\ell}$. Substitution of the expansion \eref{e:Fexp} of $F_{1\ldots\ell}$ into the cluster expansion \eref{e:G2}--\eref{e:G4} leads to
\begin{equation}
  G_{1\ldots\ell}=2^{-\ell}\sum_{a\in\mathcal{I}^\ell}g_\ell^{a}\prs_{(1,\dots,\ell)}^a,
\end{equation}
where $g_\ell^{a}$ are expressed in terms of $f_\ell^{a}$ by relations of the type
\numparts
\begin{eqnarray}
  f_2^{ab} &=& g_2^{ab}+f_1^af_1^b,\label{e:g2}\\
  f_3^{abc} &=& g_3^{abc}+f_1^af_2^{bc}
  +f_1^bf_2^{ca}+f_1^cf_2^{ab}-2f_1^af_1^bf_1^c,\label{e:g3}\\
  f_4^{abcd} &=& g_4^{abcd}+f_1^af_3^{bcd}+f_1^bf_3^{acd}+f_1^cf_3^{abd}+f_1^df_3^{abc}\nonumber\\
  &&+f_2^{ab}f_2^{cd}+f_2^{ac}f_2^{bd}+f_2^{ad}f_2^{bc}\nonumber\\
  &&-2f_1^af_1^bf_2^{cd}-2f_1^af_1^cf_2^{bd}-2f_1^af_1^df_2^{bc}-2f_1^bf_1^cf_2^{ad}\nonumber\\
  &&-2f_1^bf_1^df_2^{ac}-2f_1^cf_1^df_2^{ab}+6f_1^af_1^bf_1^cf_1^d.\label{e:g4}
\end{eqnarray}
\endnumparts
The correlation closures of orders $\ell=2,3,4$ discussed in \sref{s:closure} are obtained by setting $g_{\ell+1}=0$.

\subsection{BBGKY in terms of the expansion coefficients}
\label{s:BBGKYexpansion}
Substituting the expansions \eref{e:Fexp} into the {BBGKY} hierarchy \eref{e:spinkin}, the time evolution equations for the coefficients $f_\ell$ can be obtained. After a cumbersome but rather straightforward calculation which is reported in  \ref{s:derivation}, one finds the {BBGKY} hierarchy in terms of the coefficients $f_\ell$ to be
\begin{equation}\label{e:bbgky1}
 \frac{\hbar}{2}\partial_t f_\ell^a = v_{\ell0}^a(f_\ell) +
  \lambda v_{\ell-}^a(f_{\ell-1}) +  
 (1-\ell\lambda)v_{\ell+}^a(f_{\ell+1}),
\end{equation}
where the definition $\lambda=1/N$ has been introduced. Moreover, we defined
\numparts
\begin{eqnarray}
  v_{\ell0}^a(f_\ell)&=&-\sum_{b,c\in\mathcal{I}} h^b \sum_{i=1}^\ell \eps^{a_i b c} f_\ell^{a-a_i+c},\label{e:v-1} \\
  v_{\ell-}^a(f_{\ell-1})&=& -\sum_{b,c\in\mathcal{I}} \sum_{\scriptstyle i,j=1\atop \scriptstyle i\neq j}^\ell \eps^{a_i bc} J^{ba_j} f_{\ell-1}^{a-a_i+c-a_j},\label{e:v-2}\\
  v_{\ell+}^a(f_{\ell+1})&=&-\sum_{b,c,d\in\mathcal{I}} J^{bd} \sum_{i=1}^\ell \eps^{a_ibc}f_{\ell+1}^{a-a_i+c+d},\label{e:v-3}
\end{eqnarray}
\endnumparts
where $\varepsilon^{abc}$ is the Levi-Civita symbol defined according to the convention $\varepsilon^{xyz}=1$. With an abominable abuse of notation,
\numparts
\begin{equation}
  a-a_i+c=(a_1,\ldots,a_{i-1},c,a_{i+1},\ldots,a_\ell)
\end{equation}
in \eref{e:v-1} denotes the multi-index which is obtained from $a$ by replacing its $i$th element by $c$. Similarly,
\begin{equation}
a-a_i+d-a_j=(a_1,\ldots,a_{i-1},d,a_{i+1},\ldots,a_{j-1},a_{j+1},\ldots,a_\ell)
\end{equation}
in \eref{e:v-2} is derived from $a$ by replacing the $i$th element by $d$ and then deleting the $j$th entry of $a$, and
\begin{equation}
a-a_i+d+c=(a_1,\ldots,a_{i-1},d,a_{i+1},\ldots,a_\ell,c)
\end{equation}
\endnumparts
in \eref{e:v-3} is obtained from $a$ by replacing the $i$th element by $d$ and then appending $c$.

All the terms in $v_{\ell0}$ and $v_{\ell\pm}$ are linear in the coefficients $f_{\ell}$ and $f_{\ell\pm1}$. As inherited from the Hamiltonian, $v_{\ell0}$ contains only the local field components $h^a$, while $v_{\ell\pm}$ contain the spin interaction matrix $J$. \Eref{e:bbgky1} shows that the $\ell$th order coefficients $f_\ell^a$ are coupled to coefficients of orders $\ell\pm1$, but not to coefficients $f_n^a$ with $|n-\ell|>1$. It is this second-order recursion structure that is at the basis of the results derived in the following. 

\section{The thermodynamic Limit}
\label{s:thermodynamic}
The thermodynamic limit is a delicate issue in general, especially in the presence of long-range interactions. Recent work, mostly on classical systems, has taught us that complications increase further when one is interested in long-time asymptotics (or long-time averages) of long-range interacting systems, and their thermodynamic limit. The issue is illustrated by the time evolution of the expectation value $\langle A\rangle(t)$ in \fref{f:decay} where, for the Emch-Radin model, it was observed that relaxation to equilibrium takes place on a time scale that diverges with the system size $N$ as $N^r$, with an exponent $r=\min\{1/2,1-\alpha\}>0$. As a consequence, depending on the order in which the long-time average and the large-system limit are taken, different limit values are obtained: Performing first the long-time average%
\footnote{For finite system sizes, the long-time limit is not well defined, as the time evolution is periodic and recurrences (Loschmidt echos) occur. The period of these recurrences, however, is exceedingly long for reasonably large systems, and the excursions away from the ensemble average are short-lived. For these reasons, the time-average of the expectation value $\langle A\rangle(t)$ has a well-defined long-time limit which, for large $N$, is close to the ensemble average. When the large-system limit is taken first, these problems do not occur, the long-time limit exists and is equal to the long-time average.}
and then the large-system limit yields the equilibrium value $\langle A\rangle=0$, whereas the reverse order of the limits results in $\langle A\rangle=\langle A\rangle(0)$. More generally, instead of taking one limit after the other, one can consider any path to infinity in the ($N$,$t$) plane and take both limits simultaneously along this path. It will depend on the physical question of interest which limiting procedure and which path is the suitable one. In this section we discuss one such path which, as we shall see, is appropriate for studying the approach to thermodynamic equilibrium of the long-range interacting spin system \eref{e:Hamiltonian} and \eref{e:V12}.

As always when performing a thermodynamic limit, it is essential to identify suitably defined quantities for which this limit is well-defined and nontrivial. With this aim in mind, consider the scaling transformations
\numparts\begin{eqnarray}
  t &=& \hbar\tau \lambda^{-r} /2,\label{e:scale1}\\
  f_\ell &=& f_{\ell}'\lambda^{s\ell},\label{e:scale2}
\end{eqnarray}\endnumparts
where the parameters $r$ and $s$ are still to be determined. Substitution into \eref{e:bbgky1} and multiplication by $\lambda^{-r-s\ell}$ yields the {BBGKY} hierarchy in the scaled variables,
\begin{equation}\label{e:bbgky2}
  \partial_{\tau} f_\ell' = \lambda^{-r}v_{\ell0}(f'_\ell) +
  \lambda^{1-r-s} v_{\ell-}(f'_{\ell-1}) + 
  \lambda^{s-r}(1-\ell\lambda)v_{\ell+}(f'_{\ell+1}).
\end{equation}
The first term on the right-hand side of this equation depends, via $v_{\ell0}$, on the magnetic field components $h^a$ (i.e.\ on the on-site potential), but not on the spin-spin coupling $J$. The effect of such a constant field on the dynamics is known to be a simple spin precession, unrelated to the relaxation to equilibrium our analysis is focussing on. Moreover, as we will see later, its frequency turns out to be divergent on the time scale of equilibration.

It is therefore convenient to hide the presence of the first term on the right-hand side of \eref{e:bbgky2} by writing the BBGKY hierarchy \eref{e:spinkin} in some kind of \emph{interaction picture}. This amounts to absorbing the unitary time evolution, caused by the one-spin terms $H_i$ in the Hamiltonian \eref{e:Hamiltonian}, in the definition of a new set of reduced density operators,
\begin{equation}\label{e:iPic}
  U_{1\ldots\ell}= \exp\biggl[\frac{\rmi t}{\hbar}\sum_{i=1}^\ell H_i\biggr] F_{1\ldots\ell} \exp\biggl[-\frac{\rmi t}{\hbar}\sum_{i=1}^\ell H_i\biggr].
\end{equation}
The {BBGKY} hierarchy in the interaction picture is then given by
\begin{equation}\label{e:spinkin2}
  \fl \rmi\hbar\partial_t U_{1\ldots\ell} =
  \sum_{\scriptstyle i,j=1\atop \scriptstyle i<j}^\ell\pb{\tilde{V}_{ij}(t),U_{1\ldots\ell}}+ (N-\ell) \Tr_{\ell+1}\sum_{i=1}^{\ell}\pb{\tilde{V}_{i,\ell+1}(t),U_{1\ldots\ell+1}},
\end{equation}
where the two-body potential now is explicitly time-dependent,
\begin{equation}\label{e:V2}
  \tilde{V}_{ij}(t) =
  \exp\left[\frac{\rmi t}{\hbar} \left(H_i+H_j\right)\right]V_{ij}\exp\left[-\frac{\rmi t}{\hbar} \left(H_i+H_j\right)\right].
\end{equation}
To compute the explicit form of $\tilde{V}_{ij}$, note that Pauli operators on different lattice sites commute, and therefore the exponentials of sums in \eref{e:V2} can be factorized into a product of exponentials. Define the matrix $B(b)$ as
\begin{eqnarray}\label{e:B}
  B(b)\sigma &=& e^{-\rmi \sigma b}\sigma e^{\rmi \sigma b}\nonumber\\
  &=&\sigma\cos{\abs{2b}}
  + 2\hat{b}(\hat{b}\cdot\sigma)\sin^2{\abs{b}} 
  + (\sigma\times\hat{b})\sin{\abs{2b}},
\end{eqnarray}
with $b=(b^x,b^y,b^z)$, $\sigma=(\sigma^x,\sigma^y,\sigma^z)$, $\abs{b}=[(b^x)^2+(b^y)^2+(b^z)^2]^{1/2}$, and $\hat{b}=b/\abs{b}$. Applying \eref{e:B} to the exponential factors in \eref{e:V2} yields
\begin{equation}
\tilde{V}_{12}=-\frac{1}{N}\sum_{a,b\in\mathcal{I}} \tilde{J}^{ab}\sigma_i^a \sigma_{j}^b
\end{equation}
in a form that is equivariant to $V_{12}$ in \eref{e:V12}, with a transformed, time-dependent coupling matrix
\begin{equation}\label{e:JPic}
  \tilde{J}= B(ht/\hbar)^\trsp J B(ht/\hbar).
\end{equation}
Time appears only in the trigonometric functions in $B$, therefore $B(ht/\hbar)$ is periodic with period $T=\pi\hbar/\abs{h}$. The elements of $B(b)$ can be represented in the form
\begin{equation}
  B(b)=\left(\!\!\!\begin{array}{ccc}
  c^0+c^{xx} & c^{xy}+c^z & c^{xz}-c^y \\
  c^{xy}-c^z & c^0+c^{yy} & c^{yz}+c^x \\
  c^{xz}+c^y & c^{yz}-c^x & c^0+c^{zz} \end{array}\!\!\!\right),
\end{equation}
where $c^0=\cos{\abs{2b}}$, $c^i=\hat{b}^i\sin{\abs{2b}}$, and $c^{ij}=2\hat{b}^i\hat{b}^j\sin^2{\abs{b}}$ with $i,j\in\mathcal{I}$.

Comparing the original BBGKY hierarchy \eref{e:spinkin} to the one in the interaction picture \eref{e:spinkin2}, we observe that the single-spin term---whose diverging frequency might be bothering us in the thermodynamic limit---can be eliminated from the {BBGKY} hierarchy by passing to the interaction picture, at the expense of a time-periodic interaction potential \eref{e:V2}. Next we introduce, entirely analogous to equation \eref{e:Fexp}, an expansion of $U_{1\ldots\ell}$ in terms of coefficients $u_\ell^a$,
\begin{equation}\label{e:Uexp}
  U_{1\ldots\ell} = 2^{-\ell}\sum_{n=0}^{\ell}
  \sum_{a\in \mathcal{I}^n} u_n^a
  \sum_{p\in \setP_n(\ell)}\prs_p^a.
\end{equation}
Writing the BBGKY hierarchy in the interaction picture \eref{e:spinkin2} in terms of these expansion coefficients, we obtain
\begin{equation}\label{e:bbgky3}
  \partial_{\tau} u_\ell' = \lambda^{1-r-s}\tilde{v}_{\ell-}(u_{\ell-1}',\tau) + \lambda^{s-r}(1-\ell\lambda)\tilde{v}_{\ell+}(u_{\ell+1}',\tau).
\end{equation}
where $u_\ell=u_{\ell}'\lambda^{s\ell}$. The $\tilde{v}_{\ell\pm}(u'_{\ell\pm1},\tau)$ terms are obtained from the corresponding $v_{\ell\pm}$ terms in \eref{e:v-2} and \eref{e:v-3} by replacing $J$ with
\begin{equation}\label{e:Jtilde}
\tilde{J}(\tau)= B(\lambda^{-1/2}h \tau/2)^\trsp JB(\lambda^{-1/2}h \tau/2)
\end{equation}
as defined in \eref{e:JPic}. Inspection of \eref{e:bbgky3} suggests that the leading-order $\lambda$-dependent factors can be scaled to unity\footnote{This argument is based on the assumption that $\tilde{v}_\ell^-$ and $\tilde{v}_\ell^+$ are of order unity in $\lambda$. It will be made explicit for the special case discussed in \sref{s:case}.} by choosing $1-r-s=0$ and $s-r=0$. The solution $r=s=1/2$ determines the scaling exponents in \eref{e:scale1} and \eref{e:scale2},
\numparts\begin{eqnarray}
  t &=& \hbar\tau \lambda^{-1/2}/2,\label{e:scale1b}\\
  f_\ell &=& f_{\ell}'\lambda^{\ell/2},\label{e:scale2b}
\end{eqnarray}\endnumparts
and the BBGKY hierarchy simplifies to
\begin{equation}\label{e:bbgky4}
  \partial_{\tau} u_\ell' = \tilde{v}_{\ell-}(u_{\ell-1}',\tau) + (1-\ell\lambda)\tilde{v}_{\ell+}(u_{\ell+1}',\tau).
\end{equation}

The frequency of the modulation of $\tilde{J}(\tau)$ in \eref{e:Jtilde} has a singular $\lambda\to0$ limit, which is a hallmark of the fundamentally different time scales involved: The {\em fixed}\/ time scale of the single-spin interaction, contained in the term $v_{\ell0}(f)$ in \eref{e:bbgky1} in the ($t,N$)-plane, appears as a {\em variable}\/ frequency oscillation with a singular $\lambda=0$ limit in the parameter space ($\tau,\lambda$). Since this frequency diverges on the $\tau$-time scale we are interested in, it is reasonable to eliminate the high-frequency oscillations by applying an averaging procedure to equation \eref{e:bbgky4}. To this purpose, we apply on both sides of \eref{e:bbgky4} the finite-time averaging operator 
\begin{equation}\label{e:avgop}
  \frac{1}{T}\int_\tau^{\tau+T} d\tau',
\end{equation}
where $T=\pi\sqrt{\lambda}/|h|$ is the period of $\tilde{J}(\tau)$. For small $\lambda$, $u_\ell$ is slowly varying over a period $T$, and by making use of the definitions \eref{e:v-2} and \eref{e:v-3} we can write
\begin{equation}
\fl\eqalign{
\frac{{u'}_\ell^a(\tau+T)-{u'}_\ell^a(\tau)}{T}
\approx& -\sum_{b,c\in\mathcal{I}} \sum_{\scriptstyle i,j=1\atop \scriptstyle i\neq j}^\ell \eps^{a_i bc} \bar{J}^{ba_j} {u'}_{\ell-1}^{a-a_i+c-a_j}(\tau)\\
&-(1-\ell\lambda)\sum_{b,c,d\in\mathcal{I}} \bar{J}^{bd} \sum_{i=1}^\ell \eps^{a_ibc}  {u'}_{\ell+1}^{a-a_i+c+d}(\tau),
}
\end{equation}
where the averaged coupling matrix is defined as
\begin{equation}\label{e:Jbar}
  \bar{J} = \frac{1}{T}\int_{0}^{T} B(\theta)^\trsp J B(\theta)\rmd\theta.
\end{equation}
Performing the limit $\lambda\to0$ (which implies $T\to0$), we obtain the averaged {BBGKY} equations
\begin{equation}\label{e:bbgky5}
  \partial_{\tau} u'_\ell = \bar{v}_{\ell-}(u'_{\ell-1}) + \bar{v}_{\ell+}(u'_{\ell+1}),
\end{equation}
where $\bar{v}_{\ell\pm}$ is obtained from the corresponding $v_{\ell\pm}$ by replacing $J$ with $\bar J$. 

Comparing the original (non-averaged) equations \eref{e:bbgky2} with the averaged ones in \eref{e:bbgky5}, the outcome of the averaging procedure can be summarized as follows:
\begin{enumerate}
\item The presence of a single-spin potential $H_i$ in the Hamiltonian leads to a fast oscillating term.
\item Going to the interaction picture and averaging-out the fast oscillations, the averaged BBGKY hierarchy \eref{e:bbgky5} is formally identical to the original one \eref{e:bbgky2} {\em in the absence of a single-spin potential}, but with the original coupling matrix $J$ replaced by the averaged one $\bar J$ defined in \eref{e:Jbar}.
\item For the investigation of the long-time dynamics in the thermodynamic limit it is therefore sufficient to study the Hamiltonian \eref{e:Hamiltonian} in the absence of a single-spin potential $H_i$, thereby avoiding the problem of fast oscillations in the rescaled BBGKY hierarchy \eref{e:bbgky2}.
\item Our scaling ansatz \eref{e:scale1} and \eref{e:scale2} suggests to define the thermodynamic limit as $1/N=\lambda\to0$ at fixed rescaled time $\tau$, and this suggestion will be further investigated in \sref{s:case}. This limit corresponds to a non-trivial path to infinity in the ($t,N$) plane.
\end{enumerate}
Taking for example the special case discussed in \sref{s:closure}, $J=\mathrm{diag}(J^x,J^y,J^z)$ and $h=(0,0,h^z)$, it is straightforward to calculate the averaged coupling matrix
\begin{equation}\label{e:Javg}
  \bar{J}=\mathrm{diag}\left( (J^x+J^y)/2 , (J^x+J^y)/2, J^z \right).
\end{equation}
The above results then assert that, on the $\tau$-time scale defined in \eref{e:scale1b}, equilibration of a system with coupling $\bar{J}$ and zero magnetic field will look identical to that of a system with the original couplings and a non-zero field in $z$-direction. For the special case studied in \sref{s:case} where an exact solution (without averaging) is feasible, these considerations will be nicely confirmed and illustrated.

As a consequence of the elimination of the fast oscillations, the averaged time evolution equations \eref{e:bbgky5} are much more suitable for numerical computations, yielding more stable results. Comparing the results with and without averaging, we observe that the reduced one-spin density operator retains positivity over the period of time shown in \fref{f:diverg}. Moreover, the solution appears periodic, which is in agreement with the analytic prediction for the special case investigated in \sref{s:case}.

\section{Case study}
\label{s:case}
In this section we derive, for the choice of parameters
\begin{equation}\label{e:icrepeated}
  J=\mathrm{diag}( J^\perp,J^\perp,J^z ),\qquad h=(0,0,h^z),
\end{equation}
the time-evolution of certain coefficients $f_\ell$ in the BBGKY hierarchy \eref{e:bbgky2}. This is achieved for correlation closures of arbitrary order, and also for the full (untruncated) hierarchy. As for initial conditions, we set
\begin{equation}\label{e:ic2}
  (f_1^x,f_1^y,f_1^z)(0)=(s^x,s^y,s^z),\qquad \abs{s}\le1.
\end{equation}
The parameters \eref{e:icrepeated} and the initial conditions \eref{e:ic2} are more general than the ones used in the Emch-Radin model (where $J^\perp=0$ and the initial density operator is diagonal in the $\sigma^x$ eigenbasis, implying $f_1^y=f_1^z=0$). For the higher order coefficients we require
\begin{equation}\label{e:ic3}
  f_\ell^{a}(0)=s_\ell^{a}=0\qquad \forall\ell>1.
\end{equation}
A study of more general initial conditions and parameters will be presented elsewhere. The results we will obtain in this section are found to be in agreement with the known exact solution of the Emch-Radin model. This can be seen as an {\em a posteriori}\/ justification of the scaling assumptions \eref{e:scale1b}--\eref{e:scale2b} and of the limiting procedure proposed in \sref{s:thermodynamic}. Apart from this reassuring result, the derivation is interesting to follow as it resorts to a number of beautiful mathematical concepts, including recurrence relations, continued fractions, and orthogonal polynomials.

\subsection{Finite system size}
\label{s:specialfinite}
Unless stated otherwise, we will drop from now on the prime in the rescaled coefficients \eref{e:scale2}, writing $f$ instead of $f'$. We start by analyzing the hierarchy \eref{e:bbgky2}, and the terms $v_{\ell0}$ and $v_{\ell\pm}$ contained therein, for the above parameter values and initial conditions \eref{e:icrepeated}--\eref{e:ic3}. In the following, whenever convenient, we will label the coefficients $f_\ell^m$ by a triple $\tmb=(\tm_x,\tm_y,\tm_z)$ as introduced in \sref{s:definition}. Considering $\tmb=(0,0,\ell)$, we find $v_{\ell0}=0$ and
\begin{equation}
  \partial_{\tau} f_\ell^{(0,0,\ell)} = \ell (J^\perp-J^\perp) f_{\ell+1}^{(1,1,\ell-1)}=0,
\end{equation}
from which we can conclude that 
\begin{equation}\label{e:trivz}
  f_\ell^{(0,0,\ell)}(\tau)=f_\ell^{(0,0,\ell)}(0)=s^{(0,0,\ell)}
\end{equation}
for all $\ell$. In particular, for $\ell=1$ we have
\begin{equation}
f_1^z(\tau)=f_1^z(0)=s^z.
\end{equation}
From \eref{e:trivz}, we also note that the initial conditions \eref{e:ic3} may be generalized to nonzero $s^{(0,0,\ell)}$ at no additional expense. Next we discuss the time evolution of coefficients $f_\ell^\tmb$ for two families of superscript indices, $\tmb=(1,0,\ell-1)$ and $\tmb=(0,1,\ell-1)$,
\numparts
\begin{eqnarray}
  \fl \partial_\tau
  f_\ell^{(1,0,\ell-1)}&=&\lambda^{-1/2}h^zf_\ell^{(0,1,\ell-1)}-(\ell-1)Kf_{\ell-1}^{(0,1,\ell-2)}-(1-\lambda\ell)Kf_{\ell+1}^{(0,1,\ell)},\label{e:esp1}\\
  \fl \partial_\tau
  f_\ell^{(0,1,\ell-1)}&=&-\lambda^{-1/2}h^zf_\ell^{(1,0,\ell-1)}+(\ell-1)Kf_{\ell-1}^{(1,0,\ell-2)}+(1-\lambda\ell)Kf_{\ell+1}^{(1,0,\ell)},\label{e:esp2}
\end{eqnarray}
\endnumparts
where $K=J^z-J^\perp$. The regularity behind this set of differential equations can be captured with the help of a few definitions. Define the sequences of component labels
\numparts
\begin{eqnarray}
  \seta^x&=&(x, yz, xzz, yzzz,xzzzz,\ldots),\label{e:setax}\\
  \seta^y&=&(y, xz, yzz, xzzz,yzzzz,\ldots),\label{e:setay}
\end{eqnarray}
\endnumparts
and denote by $\seta^{x}_n$ and $\seta^y_n$ the $n$th element of the corresponding sequence, $1\le n\le N$. Furthermore, define a sequence of complex functions
\begin{equation}\label{e:u}
  u_n(\tau) = \rme^{\rmi \lambda^{-1/2}h^z \tau}\bigl[
  (-1)^{\lfloor n/2\rfloor} f_{n}^{\seta^x_n}(\tau) + \rmi 
  (-1)^{\lfloor (n-1)/2\rfloor} f_{n}^{\seta^y_n}(\tau)\bigr],
\end{equation}
where $\lfloor n/2\rfloor$ denotes the integral part of $n/2$. With these definitions, the {BBGKY} equations \eref{e:esp1} and \eref{e:esp2} can be expressed in terms of the functions $u_n$,
\begin{equation}\label{e:uk0}
 \partial_{\tau}u_n=K(n-1)u_{n-1}-K(1-n\lambda)u_{n+1},\qquad 1\le n\le N.
\end{equation}
This equation may be further simplified by another scaling of time $\tau'=\tau/K$, yielding
\begin{equation}\label{e:uk}
 \partial_{\tau}u_n=(n-1) u_{n-1}-(1-n\lambda)u_{n+1},\qquad 1\le n\le N,
\end{equation}
where we have dropped the prime from $\tau'$ for the sake of a simpler notation. Translating the initial conditions \eref{e:ic2} and \eref{e:ic3} to our new variables, we obtain
\begin{equation}\label{e:ic4}
  u(0)=(s^x+\rmi s^y,0,\ldots,0).
\end{equation}
Inspecting equation \eref{e:u}, we observe that indeed the singular term $\lambda^{-1/2}v_{\ell0}$ in \eref{e:bbgky2} results in a high-frequency oscillation superimposed on the much slower dynamics induced by the spin-spin interactions. Incorporating this oscillation into the definition of $u$ corresponds to a change of variables to the co-moving frame, i.e.\ to the interaction picture defined in \sref{s:thermodynamic}. Furthermore, \eref{e:u} justifies the assumption, mentioned in a footnote in \sref{s:thermodynamic}, that $\tilde{v}_{\ell-}$ and $\tilde{v}_{\ell+}$ are of order unity in $\lambda$. This gives support to the reasoning behind the choice of the exponents $r$ and $s$ in the scaling ansatz \eref{e:scale1}--\eref{e:scale2}. However, it also implies that, on the $\tau$-time scale, no truncation of the BBGKY hierarchy can be justified by the smallness of the parameter $\lambda$. 

Finally, the particular choices of parameter values \eref{e:icrepeated} and initial conditions \eref{e:ic2} and \eref{e:ic3} led to a simplification of the BBGKY hierarchy. The simplification becomes even more pronounced if one focusses, as we did in the preceding paragraph, onto a certain subset of the coefficients $f_\ell^a$ in the expansion \eref{e:Fexp}. For example, in order to compute expectation values of single-particle observables, it is sufficient to determine the time-evolution of $f_1^x$, $f_1^y$, and $f_1^z$. From \eref{e:uk}, we can read off that $f_1^x+\rmi f_1^y$ is determined as a solution of $N$ equations with $N$ independent variables, which is a tiny fraction of the total number of degrees of freedom $D(N)$ of the full {BBGKY} hierarchy [$D(N)\sim N^3/6$ as given by \eref{e:dofcount}]. It is important to note that this reduction of complexity is obtained without truncation of the hierarchy, but is a consequence of the decoupling of certain subsets of the expansion coefficients introduced in \eref{e:Fexp}.

The initial value problem specified by \eref{e:uk} and \eref{e:ic4} can be solved in Laplace space. The Laplace transform of a function $u$ of a real variable $\tau$ is a complex function $\Lu$ of a complex variable $z$, defined by
\begin{equation}\label{e:Lapt}
  \Lu(z)=\Lapt[f]\equiv\int_0^\infty u(\tau)\rme^{-z\tau}\rmd\tau.
\end{equation}
The inverse of the Laplace transform is defined as
\begin{equation}\label{e:iLapt}
  u(\tau)=\Lapt^{-1}[\Lu]\equiv\frac{1}{2\pi\rmi}\int_{\gamma} \Lu(z)\rme^{z\tau} \rmd z,
\end{equation}
where $\gamma$ denotes the so-called Bromwich contour in the complex $z$ plane, which runs from $-\rmi\infty$ to $+\rmi\infty$ and stays to the right of all of the poles of $\Lu(z)$. The Laplace transformation of a derivative is
\begin{equation}
\Lapt[\partial_{\tau}u(\tau)]=z\Lu(z)-u(0).
\end{equation} 
By virtue of this rule, we can write \eref{e:uk} in Laplace space,
\begin{equation}\label{e:Luk}
  z \Lu_n(z) = u_n(0) + (n-1)\Lu_{n-1}(z) -(1-n\lambda)\Lu_{n+1}(z).
\end{equation}
From equation \eref{e:Luk} with $n=1$ we obtain\footnote{We disregard the problem of zero denominators in the following, as it can be circumvented and does not cause serious problems.}
\begin{equation}
  u_1(0)/\Lu_1(z)=z+(1-\lambda)[\Lu_2(z)/\Lu_1(z)],
\end{equation}
and for $n>1$ it can be rearranged to 
\begin{equation}
  z=(n-1)[\Lu_{n-1}(z)/\Lu_n(z)-(1-n\lambda)[\Lu_{n+1}(z)/\Lu_n(z)].
\end{equation}
We can then write the ratio of subsequent $\Lu_{n}$ as
\begin{equation}\label{e:cfn}
  \frac{\Lu_{n}(z)}{\Lu_{n-1}(z)}=\frac{n-1}{z+(1-n\lambda)[\Lu_{n+1}(z)/\Lu_n(z)]}.
\end{equation}
Applying this expression iteratively, we arrive at
\begin{equation}\label{e:cfrac}
  \Lu_1(z)=(s^x+\rmi s^y)\cfra{1}{z}\cfra{\beta_1}{z}\cdots\frac{\beta_{N-1}}{z}
\end{equation}
with
\begin{equation}\label{e:beta}
  \beta_n=n(1-n\lambda),\qquad 1\le n\le N,
\end{equation}
where
\begin{equation}
  \cfra{a_1}{b_1}\cfra{a_2}{b_2}\cdots \equiv 
  \frac{a_1}{b_1 + {\displaystyle\frac{a_2}{b_2+\cdots}}}
\end{equation}
is a standard notation for continued fractions.

From $\Lu_1$, the coefficients $\Lu_n$ with $n>1$ are obtained via the equation \eref{e:Luk}. For a finite number $N=1/\lambda$ of spins, the solution \eref{e:cfrac} contains a finite number of terms, terminating at $\beta_{N}=N(1-N\lambda)=0$. Hence we can write
\begin{equation}\label{e:ratio}
  \Lu_1(z)= (s^x+\rmi s^y)\frac{A_N(z)}{B_N(z)},
\end{equation}
where $A_N(z)$ and $B_N(z)$ are polynomials in $z$ of degree $N-1$ and $N$, respectively.

From $\Lu_1$ we reconstruct the coefficients $f_1^{x}$ and $f_1^y$ in the time domain via the inverse Laplace transform \eref{e:iLapt},
\begin{equation}\label{e:backt}
  f_1^x(\tau)+\rmi f_1^y(\tau)=\rme^{-\rmi\lambda^{-1/2}h^z\tau}\Lapt^{-1}[\Lu_1(z)](\tau).
\end{equation}
Properties of \eref{e:backt} depend crucially on the properties of $\Lapt^{-1}[\Lu_1(z)]$, and therefore on the structure of the singularities of $\Lu_1(z)$. Owing to the form of the expression \eref{e:ratio}, these singularities are poles, originating from the zeros of $B_{N}(z)$. We shall study these zeros in the following and discuss the effect of correlation closures of various orders $\ell$ on the solutions. For this purpose, consider what is called the $\ell$th {\em convergent}\/ of the continued fraction \eref{e:cfrac},
\begin{equation}\label{e:cfractrunc}
\Lu_1^{(\ell)}(z)=(s^x+\rmi s^y)\cfra{1}{z}\cfra{\beta_1}{z}\cdots\frac{\beta_{\ell-1}}{z},
\end{equation}
defined as the truncation of \eref{e:cfrac} after $\ell$ terms. The same $\Lu_1^{(\ell)}$ could also be obtained as a solution of a truncated version of the recurrence relation \eref{e:Luk} where $\Lu_{\ell+k}=0$ for all $1\le k\le N-\ell$. For this reason, the convergent \eref{e:cfractrunc} is closely related to the $\ell$th order correlation closure of the BBGKY hierarchy, as given by \eref{e:g2}--\eref{e:g4} with $g_{\ell+k}=0$. This can be seen by applying the correlation closure condition to the coefficients $f_{\ell+1}^{\seta^x}$, $f_{\ell+1}^{\seta^y}$ with component indices from the sets \eref{e:setax} or \eref{e:setay}: Because of the $(n-1)$-fold occurrence of $z$-indices in the elements $\seta^x_n$ and $\seta^y_n$, the cluster expansions \eref{e:g2}--\eref{e:g4} are particularly simple. The coefficient $f_4^{\seta^y_4}$ for example can be written as 
\begin{equation}\label{e:f4example}
\fl f_4^{\seta^y_4}=f_4^{xzzz}=f_1^x\left[f_3^{zzz}-6f_1^zf_2^{zz}+6(f_1^z)^3\right]+3f_2^{xz}\left[f_2^{zz}-2(f_1^z)^2\right]+3f_3^{xzz}f_1^z+g_4^{xzzz}.
\end{equation}
In third order correlation closure, i.e.\ setting $g_4^{xzzz}=0$, the right-hand side of \eref{e:f4example} consists of only two kinds of coefficients $f_n^a$: either with $a\in \seta^x \cup \seta^y$, or with $a=z\ldots z$. The latter ones, as we have shown in \eref{e:trivz}, do not vary in time, and $f_4^{xzzz}$ is therefore linear in the time-dependent coefficients $f_n^{\seta^x}$, $f_n^{\seta^y}$. Translating this observation via definition \eref{e:u} into the interaction picture, we find the closure condition
\begin{equation}
u_4(\tau)=\sum_{n=1}^3 c_n u_n(\tau),
\end{equation}
where the $c_n$ are determined by the time-independent coefficients $f_k^\tmb$ with $\tmb=(0,0,k)$. The same kind of pattern emerges also for correlation closures of arbitrary order $\ell$, where
\begin{equation}\label{e:cc}
  u_{\ell+1}(\tau)=\sum_{n=1}^\ell c_n u_{n}(\tau)
\end{equation}
is again linear in the time-dependent coefficients $u_{n}(\tau)$. In the special case of initial conditions $s^{(0,0,k)}=f^{(0,0,k)}=0$, we have $c_n=0$ and the $\ell$th order correlation closure is equivalent to setting $u_n=0$ for $n>\ell$. This condition translates into the condition $\Lu_{\ell+k}=0$ for the coefficients in Laplace space, and \eref{e:cfractrunc} is therefore the solution of the $\ell$th order correlation closure for initial conditions given by \eref{e:ic3}. In case of more general initial conditions with $s^{(0,0,k)}\neq0$, the correlation closure amounts to a linear relation between $u_{\ell+1}$ and all the lower order coefficients.

Having understood the relation between correlation closures \eref{e:correlationclosure} and convergents \eref{e:cfractrunc}, we proceed with the analysis of the latter. Similar to the full continued fraction in \eref{e:ratio}, also the convergent can be written as a ratio of polynomials $A_{\ell}$ and $B_{\ell}$ of degrees $\ell-1$, respectively $\ell$,
\begin{equation}\label{e:ratio_ell}
  \Lu_1^{(\ell)}(z)= (s^x+\rmi s^y)\frac{A_\ell(z)}{B_\ell(z)}.
\end{equation}
From \eref{e:cfractrunc}, Wallis-type \cite{Chihara} second-order recursion relations for $B_n$ and $A_n$ can be derived,
\numparts
\begin{eqnarray}
  zA_{n}(z)&=& \beta_{n}A_{n-1}(z) - A_{n+1}(z),\label{e:recursionA}\\
  zB_{n}(z)&=& \beta_n B_{n-1}(z) - B_{n+1}(z),\label{e:recursionB}
\end{eqnarray}
\endnumparts
where $\beta_n$ is defined as in \eref{e:beta}. Although of the same form, the two recursion relations \eref{e:recursionA} and \eref{e:recursionB} have different initial conditions, $B_{-1}=0$, $B_0=1$, and $A_0=0$, $A_1=-1$. Anticipating that the zeros of $B_n$ and $A_n$ are imaginary, we make a coordinate transformation $x=-\rmi z$ that brings $A_n$ and $B_n$ into the conventional real form. In these new variables, the convergent can be written as
\begin{equation}\label{e:ratio2}
  \Lu_1^{(\ell)}(\rmi x)= -\rmi (s^x+\rmi s^y)
  \frac{p^{(1)}_{\ell-1}(x)}{p^{(0)}_\ell(x)},
\end{equation}
with polynomials
\begin{equation}
p^{(1)}_n(x) = -\rmi^nA_{n+1}(\rmi x),\qquad p^{(0)}_n(x)= \rmi^nB_n(\rmi x).
\end{equation}
From \eref{e:recursionA} and \eref{e:recursionB}, recursion relations for the new polynomials are obtained,
\begin{equation}\label{e:wallis}
  xp^{(j)}_{n}(x)=\beta_{n+j} p^{(j)}_{n-1}(x) + p^{(j)}_{n+1}(x),\qquad j=0,1,
\end{equation}
with initial conditions $p^{(j)}_{-1}=0$ and $p^{(j)}_0=1$. It follows from Favard's theorem \cite{Chihara} that the sequences of polynomials $\{p^{(j)}_{n}\}_{n=0}^{N-j}$ generated by \eref{e:wallis} are positive definite and orthogonal, in the sense that there exists a nonnegative weight function $w^{(j)}_N(x)$ such that, for any  $0\le n< m \le N-j$, 
\begin{equation}\label{e:orthogonal}
  \int_{-\infty}^{\infty} p^{(j)}_n(x) p^{(j)}_m(x) w^{(j)}_N(x) \rmd x = 0.
\end{equation}
It follows from the positive definiteness and orthogonality of the $p^{(j)}_n$ that all $\ell$ zeros $x^j_{\ell k}$ of $p^{(j)}_\ell(x)$ are real, simple, and isolated. Most importantly, the following decomposition formula holds, 
\begin{equation}\label{e:residue}
  \frac{p^{(1)}_{\ell-1}(x)}{p^{(0)}_{\ell}(x)}=\sum_{k=1}^{\ell}\frac{a_{\ell k}}{x-x_{\ell k}},\qquad 
  a_{\ell k}=\frac{p^{(1)}_{\ell-1}(x_{\ell k})}{{p'}^{(0)}_{\ell}(x_{\ell k})},
\end{equation}
where the prime denotes a derivative. 

\Eref{e:residue} can be used to perform the integral in the inverse Laplace transform \eref{e:iLapt}. Since we have to undo the coordinate transform $z=\rmi x$, the Bromwich integration contour $\gamma$ in the complex $z$-plane has to be modified to $\gamma'$ in the complex $x$-plane, running parallel to the real axis and below all the zeros of $p^{(0)}_\ell (x)$, i.e.\ below the real axis. The exponential $\rme^{\rmi x\tau}$ in \eref{e:iLapt} introduces a damping factor if $\Im x>0$, and the integration path $\gamma'$ can therefore be closed in the upper half plane by a path $\gamma'$  without changing the value of the integral. As a result, the path encloses all $\ell$ zeros of $p^{(0)}_\ell(x)$ and we can apply the residue theorem,
\begin{equation}\label{e:residue1}
\eqalign{
  u_1^{(\ell )}(\tau)&=\frac{-\rmi (s^x+\rmi s^y)}{2\pi}
  \int_{\gamma'}\frac{p^{(1)}_{\ell -1}(x)e^{\rmi x\tau}}{p^{(0)}_\ell (x)}\rmd x\\
  &=(s^x+\rmi s^y) \sum_{j=1}^\ell \mathrm{Res}_{x=x_{\ell j}}
  \left(\frac{p^{(1)}_{\ell -1}(x)}{p^{(0)}_{\ell }(x)}\right)\rme^{\rmi x_{\ell j}\tau}\\
  &=(s^x+\rmi s^y) \sum_{j=1}^{\ell}a_{\ell j}\rme^{\rmi x_{\ell j}\tau}.
}
\end{equation}
By inspection of \eref{e:wallis}, it is not too difficult to observe that the polynomials generated by this recursion relation are alternatingly odd and even functions. Their zeros are therefore distributed symmetrically around the origin. Sorting the zeros in increasing order, \eref{e:residue1} can be simplified to
\begin{equation}\label{e:residue2}
  u_1^{(\ell )}(\tau)= 2(s^x+\rmi s^y) \sum_{j=1}^{\ell /2}a_{\ell j}\cos{(x_{\ell j}\tau)},
\end{equation}
where we have assumed that $\ell$ is even. (If $\ell$ is odd, a constant term must be added to the sum.)

\begin{figure}\center
  \includegraphics[width=0.55\textwidth]{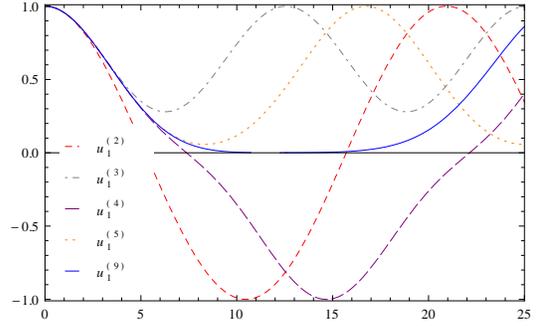}
  \hspace{3mm}
  \caption{\label{f:4} For parameter values $h=0$, $J=\mathrm{diag}(0,0,1)$, and $N=1/\lambda=10$, the real part $\Re{(u^{(\ell)}_1)}=f_1^x$ is plotted as a function of rescaled time $\tau$ for various orders $\ell=2,3,4,5,9$ of the correlation closure. Recurrent behavior is evident from the plot, both for the full solution and for the truncations of various orders.}
\end{figure}

\Eref{e:residue2} is the main result of this section. Since it consists of a finite sum of oscillatory terms, the solution $u_1^{(\ell)}(\tau)$ is either periodic or quasi-periodic in time. For example, by setting $\lambda=0$ in the polynomials $p^{(j)}_{\ell-j}$ with $\ell=2,3,4$, we find
\numparts
\begin{eqnarray}\label{e:qperiodic}
  u_1^{(2)}(\tau)&=&(s^x+\rmi s^y)\cos{\tau},\\
  u_1^{(3)}(\tau)&=&\frac{(s^x+\rmi s^y)}{3}\left[2+\cos\left(\sqrt{3/2}\tau\right)\right],\\
  u_1^{(4)}(\tau)&=&\frac{(s^x+\rmi s^y)}{2\sqrt{6}}\biggl[\left(\sqrt{6}+2\right)\cos\left({\sqrt{3-\sqrt{6}}\tau}\right) \nonumber \\
                && + \left(\sqrt{6}-2\right)\cos\left({\sqrt{3+\sqrt{6}}\tau}\right)\biggr].
\end{eqnarray}
\endnumparts
The behavior of the solutions in \eref{e:residue2} is illustrated in \fref{f:4} for various orders $\ell$ of the truncation of the continued fraction \eref{e:cfractrunc}. Initial conditions satisfying $s^z=0$ were chosen, and therefore the truncations correspond to $\ell$th order correlation closures. The recurrences, or Loschmidt echos, of the solutions, expected from the result in \eref{e:residue2}, are evident in the plots, both for the full solution for $N=10$ spins as well as for the truncations at orders $\ell<N$.

\subsection{Thermodynamic limit}
\label{s:speciallimit}
Having observed that, for any finite system size and/or finite order correlation closure, the dynamics is periodic or quasi-periodic, we next want to investigate the effect of the thermodynamic limit $\lambda\to0$. The reader is reminded that this limit, due to the $\lambda$-scaling of time \eref{e:scale1}, corresponds to simultaneously taking long-time and large-system limits along a nontrivial path in the $(t, N)$ plane. For the special case studied in this \sref{s:case}, this limit can be performed within the framework developed in \sref{s:specialfinite} by studying $p^{(1)}_{N-1}/p^{(0)}_{N}$ in the limit $\lambda\to0$. Here, however, we prefer to take a shortcut and investigate the $\lambda=0$ case directly in the time domain. Indeed, setting $\lambda=0$ in \eref{e:uk}, we obtain
\begin{equation}\label{e:uthermo}
  u_{n+1}(\tau,0)=(n-1) u_{n-1}(\tau,0)-\partial_\tau u_n(\tau,0),
\end{equation}
which is solved by 
\begin{equation}\label{e:solvd!}
  u_n(\tau,0)= (s^x+\rmi s^y) \tau^{n-1} \rme^{-\tau^2/2}.
\end{equation}
Translating this result to the original coefficients $f_{n}^{\seta^x_n}$ by \eref{e:u}, we encounter a term $\rme^{-\rmi\lambda^{-1/2}h^z\tau}$, oscillating at infinite frequency in the $\lambda\to0$ limit. This oscillation is irrelevant for the relaxation to equilibrium, as it happens on a fundamentally different timescale. Therefore we will disregard this oscillating term, obtaining
\begin{equation}\label{e:solvd2}
  f_1^x(\tau,0)+\rmi f_1^y(\tau,0) = (s^x+\rmi s^y) \rme^{-\tau^2/2}
\end{equation}
for $n=1$.

The solutions \eref{e:solvd!} or \eref{e:solvd2} display Gaussian (superexponential) relaxation as a function of rescaled time $\tau$, and no recurrent behavior is observed. This is in stark contrast to the periodic or quasi-periodic dynamics \eref{e:qperiodic} we found, as illustrated in \fref{f:4}, for finite system sizes. Moreover, the result in \eref{e:solvd2} is in exact agreement with the $N\to\infty$ limit of the Emch-Radin model with exponent $\alpha=0$ as reported in equation (32) of \cite{KastnerCEJP}.

\section{Discussion of the results}
\label{s:discussion}
The special case treated in \sref{s:case} illustrates and confirms some of the aspects that have been discussed in sections \ref{s:closure} and \ref{s:thermodynamic} in greater generality, and it also can substantiate some of assumptions made. In particular, we would like to discuss the following aspects.

\subsection{Symmetry-induced causal relations}\label{s:causal}
We have seen that, for the special case of \sref{s:case}, not all of the expansion coefficients $u_\ell^a$ (or $f_\ell^a$) influence the time evolution of each other. Instead, we observed that the time evolution of, say, $f_1^x$ is affected by only a small subset of the other coefficients. This decoupling, which is a property of the structure of {BBGKY} hierarchy of equations, is particularly pronounced in the case discussed in \sref{s:case} but, as will be shown in a forthcoming paper \cite{rytis2012}, similar (but more involved) relations hold true for general parameter values of the model.

\subsection{Shortcomings of the correlation closure}\label{s:shortcomings}
For the special values of parameters and initial conditions of \sref{s:case}, we found that, for any $\ell\geqslant1$, the correlation closure of order $\ell$ [as defined in \eref{e:correlationclosure}] gives rise to periodic or quasi-periodic behavior. There are, of course, many other types of closures of the BBGKY hierarchy one might consider, and the question is whether others might be more suitable for studying the relaxation to equilibrium we are interested in.

One way to improve on the correlation closure might consist in taking into consideration the symmetries imposed by the {BBGKY} hierarchy as mentioned in \sref{s:causal}. Since symmetries are respected by the exact dynamics, one may hope that a closure which takes into account this structure might be superior to the one that does not. Preliminary numerical studies of ours seem to be in favor of this conjecture. However, it is clear from these results that for our main objective, i.e.\ the study of relaxation to equilibrium, such a modified closure does at best lead to marginal improvements and we will hence not pursue this aspect further in the present article.

A kinetic theory appropriate for studying relaxation to equilibrium is expected to require a more complicated closure relation in the form of a collision integral. Such a theory will be developed in a forthcoming paper.

\subsection{Thermodynamic limit and the definition of rescaled time $\tau$}
A crucial step for obtaining a nontrivial long-time asymptotics was the definition of a suitable kind of thermodynamic limit as elaborated on in \sref{s:thermodynamic}. This limit reflects in the choice of the exponent $r$ when defining the rescaled time variable $\tau$ in \eref{e:scale1} \footnote{Without rescaling of time, the time evolution of the variables $u_\ell$ in the interaction picture would just be constant in the thermodynamic limit, with no sign of relaxation to equilibrium \cite{KastnerCEJP}.}. In the general considerations in \sref{s:thermodynamic}, the choice $r=1/2$ was based on the scaling properties in $\lambda$ of the BBGKY hierarchy \eref{e:bbgky3} in the coefficient expansion. With this choice, and for the special case of \sref{s:case}, we were able to derive, in \eref{e:solvd2}, the exact Gaussian relaxation to equilibrium, which gives support not only to the choice $r=1/2$, but also to the guiding principle based on the scaling properties of \eref{e:bbgky3}.

\section{Conclusions}\label{s:conclusions}

We have studied, via a BBGKY-type approach, the time evolution of $\ell$-spin reduced density operators $F_{1\dots\ell}$ for Heisenberg-type quantum spin models with Curie Weiss-type long-range interactions. Our analysis is based on a particular expansion of $F_{1\dots\ell}$ in terms of coefficients $f_\ell^a$, as introduced in \eref{e:Fexp}, which casts the BBGKY hierarchy into the form of a second-order recursion \eref{e:bbgky1}.

Originally our study was motivated by the observation of quasi-stationary behavior in the long-range Emch-Radin model, i.e.\ the fact that relaxation to equilibrium takes place on a time scale which diverges with the system size. As a consequence of this diverging time scale, for studying the long-time asymptotics of quantum spin models in the thermodynamic limit it is therefore necessary to consider a suitably defined thermodynamic limit where, with increasing system size $N$, time is scaled appropriately such that nontrivial dynamics can be observed. Remarkably, the structure of the BBGKY hierarchy \eref{e:bbgky3} when expressed in terms of the expansion coefficients $f_\ell^a$ suggests a definition of rescaled time $\tau=2t\lambda^r /\hbar$ which, as turns out, leads to a nontrivial exact result and reproduces the Gaussian relaxation \eref{e:solvd!} of the Emch-Radin model.

When dealing with the BBGKY hierarchy \eref{e:bbgky3} in rescaled time, we noticed that the on-site potential (or single-spin potential) $H_i$ in the Hamiltonian \eref{e:Hamiltonian} gives rise to an oscillating term whose frequency diverges on the $\tau$-time scale in the thermodynamic limit. Since this time scale of oscillation is strongly (infinitely) separated from the time scale of relaxation to equilibrium, an averaging procedure is introduced to eliminate the high-frequency dynamics. The averaged BBGKY hierarchy \eref{e:bbgky5} leads to a well-defined $\lambda\to0$ limit of the hierarchy, but also to a significant improvement of numerical results as shown in \fref{f:diverg}.

A general BBGKY hierarchy cannot be solved exactly, as this would correspond to an exact solution of the full $N$-spin problem. The problem becomes more tractable by truncating the hierarchy. The resulting set of equations is ill-defined, but it can be turned into a well-defined one by postulating a closure condition. For the long-range spin models discussed in the present work, we have defined what we call the correlation closure of $\ell$th order in \eref{e:correlationclosure}, and the effect of these closures on the long-time dynamics has been discussed. For the special parameter values and initial conditions considered in \sref{s:case}, we have shown analytically that closing the BBGKY hierarchy by neglecting $\ell$-spin correlations does never lead to equilibration, but gives rise to quasi-periodic time evolution with at most $\ell/2$ independent frequencies, as is evident from \eref{e:residue2}. We must therefore conclude that, in order to construct a kinetic theory appropriate for studying relaxation to equilibrium, a more complicated closure relation, presumably in the form of a collision integral, will be needed. Once such a theory is properly benchmarked against the exact results available for the Emch-Radin model, it should allow us to study the approach to equilibrium in non-integrable generalizations of the Emch-Radin model, or non-integrable long-range variants of the Heisenberg model for which exact solutions do not exist.

In addition to providing a tool for investigating the dynamics of quantum spin models, our results also shed light on more general features regarding the role of closure conditions when truncating the BBGKY hierarchy: The particularly simple structure of spin-$1/2$ lattice models facilitates analytic calculations beyond what can be achieved in continuum systems, and an understanding of the effect of approximation schemes is easier to attain. We tend to believe that, at least on a qualitative level, several of our observations apply not only to spin-$1/2$ systems, but should hold more generally for closed quantum systems on finite-dimensional Hilbert spaces.

We want to conclude with a comment on an interesting related work by Sciolla and Biroli \cite{SciollaBiroli11} which also deals with the dynamics of Curie-Weiss-type (or completely connected) quantum systems, i.e.\ with Hamiltonians of type \eref{e:Hamiltonian}, but with arbitrary $H_i$ and $V_{ij}$. Their analysis yields exact analytical results in the thermodynamic limit for a much larger class of Hamiltonians, but is more restrictive with respect to initial conditions, allowing only wave packets whose width shrinks to zero in the thermodynamic limit. Moreover, their analysis is bound to fail on the rescaled time scales $\tau$ studied in the present work, and is therefore unsuitable for investigating thermalization. In a future work we will study thermalization in rescaled time for more general models by approximative methods, and the results by Sciolla and Biroli should prove useful for benchmarking the short-time dynamical behavior. 

\appendix

\section{Derivation of equation \eref{e:bbgky1}}
\label{s:derivation}
In the course of the proof, we will refer to the following elementary identities for Pauli operators on lattice sites $i=1,2$. 
\begin{eqnarray}
  \pb{\sigma^a_i,\sigma^b_i} 
  &=&  2\rmi \sum_c \eps^{abc}\sigma^c_i,\label{e:com1} \\
 \pb{\sigma_1^a\sigma_2^u,\sigma_1^b\sigma_2^v} 
 &=& 2\rmi \sum_c \left(\delta^{uv}\eps^{abc}\sigma_1^c+ \delta^{ab}
  \eps^{uvc}\sigma_2^c\right),\label{e:com2} \\
  \Tr_2\pb{\sigma_1^a\sigma_2^u,\sigma_1^b\sigma_2^v} 
  &=& 4\rmi \delta^{uv}\sum_c \eps^{abc}\sigma_1^c,\label{e:com3}
\end{eqnarray}
with component (superscript) indices $a,b,c,u,v\in \mathcal{I}$. We use the notation $\delta^{ab}$ for the Kronecker tensor, and $\eps^{abc}$ for the Levi-Civita tensor (with $\eps^{xyz}=1$). In the identities involving Pauli operators on different lattice sites, the identity operator is implied on each single-spin Hilbert space ${\mathcal H}_i$ not acted upon by any of the Pauli operators [as on the right-hand side of \eref{e:com2}].

The starting point of the proof is the BBGKY hierarchy in the form \eref{e:spinkin} with a fixed value of $\ell$. Inserting the expansion \eref{e:Fexp} into the left-hand side of \eref{e:spinkin}, we obtain
\begin{equation}\label{e:lhs}
  2^{-\ell}\rmi\hbar\sum_{n=0}^{\ell}
  \sum_{a\in\mathcal{I}^n} \partial_t f_n^a
  \sum_{p\in \setP_n(\ell)}\bm\sigma_{p}^{a},
\end{equation}
where $\mathcal{I}=\{x,y,x\}$ denotes the set of component indices. Since the operators $\bm\sigma_{p}^{a}$ are linearly independent among each other, we can separately equate their coefficients in \eref{e:lhs} to those which result from expanding also the right-hand side of \eref{e:spinkin} in terms of \eref{e:Fexp}. We will in the following set $n=\ell$, as this will be sufficient in order to obtain time evolution equations for all coefficients $f_n^a$. Since $\setP_\ell(\ell)=\{(1,\dots,\ell)\}$ contains only a single element, the sum over $p$ in \eref{e:lhs} consists only of one term, $\bm\sigma_{(1,\dots,\ell)}^{a}$. The rest of this appendix will therefore be devoted to computing, from an expansion of the right-hand side of \eref{e:spinkin}, all the terms proportional to $\bm\sigma_{(1,\dots,\ell)}^{a}$, and then equate them to
\begin{equation}\label{e:lhs_n=l}
  2^{-\ell}\rmi\hbar \partial_t f_\ell^a \prs^{a}_{(1,\dots,\ell)}
\end{equation}
in order to obtain the time evolution equation of $f_\ell^a$. We will discuss the three terms on the right-hand side of \eref{e:spinkin} separately, showing that 
\begin{equation}\label{e:schematic}
  \eref{e:lhs_n=l} = \eref{e:term1} + \eref{e:term2} + \eref{e:term3}.
\end{equation}

We start by inserting the expansion \eref{e:Fexp} into the first term on the right-hand side of \eref{e:spinkin}, yielding
\begin{equation}\label{e:1term_exp}
\sum_{i=1}^\ell\left[H_i,F_{1\dots\ell}\right] =
-2^{-\ell}\sum_{b\in\mathcal{I}} h^b \sum_{i,n=1}^\ell \sum_{a\in\mathcal{I}^n} f_n^a \sum_{p\in \setP_n(\ell)} \bigl[\sigma_i^b,\prs_p^{a}\bigr].
\end{equation}
For the commutator in \eref{e:1term_exp} we find
\begin{equation}\label{e:c1}
\fl \bigl[\sigma_i^b,\prs_p^{a}\bigr] = \sum_{j=1}^n \delta_{i,p_j}\prs^{a-a_j}_{p-p_j} \pb{\sigma_i^b,\sigma_i^{a_j}}= -2\rmi\sum_{j=1}^n \delta_{i,p_j} \sum_{c\in\mathcal{I}} \eps^{a_j b c} \prs^{a-a_j+c}_{p},
\end{equation}
where \eref{e:com1} has been used. Here, $p-p_j$ and $a-a_j$ denote the sequences obtained from $p$ and $a$ by deleting their $j$th elements, and
\begin{equation}\label{e:tildea}
  a-a_j+c=(a_1,\ldots,a_{j-1},
  \underbrace{c}_{\hspace{-10mm}\displaystyle j\mbox{th element}\hspace{-10mm}},a_{j+1},\ldots,a_n)
\end{equation}
is the sequence where the $j$th element has been replaced by $c$. We observe that, for a given $p=(p_1,\dots,p_n)$ with $n$ elements, the commutator in \eref{e:c1} is again a product of $n$ Pauli operators acting on different lattice sites. Since, as explained above, we can restrict our attention to the terms proportional to a fixed $\bm\sigma_{(1,\dots,\ell)}^{a}$, it is sufficient to consider the terms with $n=\ell$ in \eref{e:1term_exp},
\begin{equation}\label{e:1term_2}
2^{1-\ell}\rmi\sum_{b,c\in\mathcal{I}} h^b \sum_{i,j=1}^\ell \sum_{p\in \setP_\ell(\ell)} \delta_{i,p_j} \sum_{a\in\mathcal{I}^\ell} f_\ell^a \eps^{a_j b c} \prs^{a-a_j+c}_{p},
\end{equation}
where the expression \eref{e:c1} for the commutator has been inserted. Since $\setP_\ell(\ell)=\{(1,\dots,\ell)\}$ consists of only a single element, the sum over $p$ disappears. Moreover, for the same reason, we have $\delta_{i,p_j}=\delta_{i,j}$, which allows us to execute the sum over $j$, yielding
\begin{equation}\label{e:1term_3}
2^{1-\ell}\rmi\sum_{b,c\in\mathcal{I}} h^b \sum_{a\in\mathcal{I}^\ell} f_\ell^a \sum_{i=1}^\ell \eps^{a_i b c} \prs^{a-a_i+c}_{(1,\dots,\ell)}.
\end{equation}
Swapping names of the summation indices $a_i$ and $c$ and making use of the cyclic property of the Levi-Civita tensor, we can rewrite \eref{e:1term_3} as
\begin{equation}\label{e:1term_4}
-2^{1-\ell}\rmi \sum_{a\in\mathcal{I}^\ell} \prs^{a}_{(1,\dots,\ell)}\sum_{b,c\in\mathcal{I}} h^b \sum_{i=1}^\ell \eps^{a_i b c} f_\ell^{a-a_i+c}.
\end{equation}
Owing to the mutual independence of the $\prs^{a}_{p}$, we now can pick, for a given $a$, the term
\begin{equation}\label{e:term1}
-2^{1-\ell}\rmi \prs^{a}_{(1,\dots,\ell)} \sum_{b,c\in\mathcal{I}} h^b \sum_{i=1}^\ell \eps^{a_i b c} f_\ell^{a-a_i+c}
\end{equation}
which gives the first contribution to the time evolution equation \eref{e:schematic} for the coefficient $f_\ell^a$.

To deal with the second term on the right-hand side of \eref{e:spinkin}, we again replace the $\ell$-spin reduced density operator $F_{1\dots\ell}$ by the expansion \eref{e:Fexp}, obtaining
\begin{equation}\label{e:2term_exp}
\fl \sum_{j<i=1}^\ell\pb{V_{ij},F_{1\ldots\ell}} = -\sum_{b,c\in\mathcal{I}} \frac{J^{bc}}{2^{\ell+1} N} \sum_{\scriptstyle i,j=1\atop \scriptstyle i\neq j}^\ell \sum_{n=1}^\ell \sum_{a\in\mathcal{I}^n} f_n^a \sum_{p\in \setP_n(\ell)} \bigl[\sigma_i^b \sigma_j^c,\prs_p^{a}\bigr].
\end{equation}
Under the condition $i\neq j$, we can write
\begin{equation}\label{e:d_com}
\fl \eqalign{
\bigl[\sigma_i^b \sigma_j^c,\prs_p^{a}\bigr] = &\sum_{k,l=1}^n \delta_{i,p_k}\delta_{j,p_l} \prs_{p-p_k-p_l}^{a-a_k-a_l} \bigl[\sigma_i^b \sigma_j^c,\sigma_i^{a_k}\sigma_i^{a_l}\bigr]\\
&+\sum_{k=1}^n \left(\delta_{i,p_k}\delta_{j\notin p}\sigma_j^c \bigl[\sigma_i^b,\sigma_i^{a_k}\bigr] + \delta_{j,p_k}\delta_{i\notin p}\sigma_i^c \bigl[\sigma_j^b,\sigma_j^{a_k}\bigr]\right) \prs_{p-p_k}^{a-a_k}
}
\end{equation}
for the commutator in \eref{e:2term_exp}, where we have used the short-hand notation $\delta_{j\notin p}=\prod_{m=1}^n\left(1-\delta_{j,p_m}\right)$. The first term on the right-hand side accounts for the cases where both, $i$ and $j$, have a counterpart in $\prs_p^{a}$, the second term for when only one of them has. (If neither $i$ nor $j$ have a counterpart in $\prs_p^{a}$ then the commutator is zero.) Applying \eref{e:com2} to the commutator in the first sum in \eref{e:d_com}, all summands will consist of products of $n-1$ Pauli operators. Since $n\leqslant\ell$, none of these terms can give a contribution to \eref{e:2term_exp} which is proportional to $\prs_{(1,\dots,\ell)}^{a}$, and the first sum in \eref{e:d_com} can therefore be neglected. Applying \eref{e:com1} to the commutator in the second sum in \eref{e:d_com}, however, results in products of $n+1$ Pauli operators. Hence, inserting \eref{e:d_com} into \eref{e:2term_exp}, we can restrict the summation to the $n=\ell-1$ terms, as only those may lead to contributions proportional to $\prs_{(1,\dots,\ell)}^{a}$,
\begin{equation}\label{e:2term2}
\eqalign{
\sum_{b,c\in\mathcal{I}} \frac{J^{bc}}{2^{\ell} N} \sum_{a\in\mathcal{I}^{\ell-1}} f_{\ell-1}^a \sum_{p\in \setP_{\ell-1}(\ell)} \sum_{\scriptstyle i,j=1\atop \scriptstyle i\neq j}^\ell \sum_{k=1}^{\ell-1} \prs_{p-p_k}^{a-a_k}\\
\qquad\times\left(\delta_{i,p_k}\delta_{j\notin p}\sigma_j^c \bigl[\sigma_i^b,\sigma_i^{a_k}\bigr] + \delta_{j,p_k}\delta_{i\notin p}\sigma_i^c \bigl[\sigma_j^b,\sigma_j^{a_k}\bigr]\right).
}
\end{equation}
The terms in the round brackets in \eref{e:2term2} are symmetric to each other under the exchange of $i\leftrightarrow j$, and their contributions to the overall sum are therefore identical. Hence we can write
\begin{equation}\label{e:2term3}
\fl \sum_{b,c,d\in\mathcal{I}} \frac{\rmi J^{bc}}{2^{\ell-1} N} \sum_{a\in\mathcal{I}^{\ell-1}} f_{\ell-1}^a \sum_{p\in \setP_{\ell-1}(\ell)} \sum_{\scriptstyle i,j=1\atop \scriptstyle i\neq j}^\ell \sum_{k=1}^{\ell-1} \delta_{i,p_k}\delta_{j\notin p}\eps^{a_k bd} \prs_{p+j}^{a-a_k+d+c},
\end{equation}
where \eref{e:com1} has been applied to the commutators in \eref{e:2term2}. The set $\setP_{\ell-1}(\ell)$ consists of all sequences where precisely one of the numbers $1,\dots,\ell$ is omitted. For example, for $\ell=4$, we have $\setP_{3}(4)=\{(1,2,3),(1,2,4),(1,3,4),(2,3,4)\}$. Due to the constraint $\delta_{j\notin p}$ in \eref{e:2term3}, $j$ has to equal this omitted number for all non-vanishing contributions to the sum. We can therefore combine every $p\in \setP_{\ell-1}(\ell)$ together with $j\notin p$ into a new summation index $p'\in\setP_{\ell}(\ell)$, and then use the sum over $j$ to count through the previously omitted elements of the sequence. With these new summation indices, we can rewrite \eref{e:2term3} in the form
\begin{equation}\label{e:2term4}
 \frac{2^{1-\ell}\rmi}{N} \sum_{b,d\in\mathcal{I}} \sum_{\scriptstyle i,j=1\atop \scriptstyle i\neq j}^\ell \sum_{a\in\mathcal{I}^{\ell}} \eps^{a_i bd} J^{ba_j} f_{\ell-1}^{a-a_j} \sum_{p'\in\setP_{\ell}(\ell)} \prs_{p'}^{a-a_i+d},
\end{equation}
where the former component index $c$ has now been included into the multi-index $a$. Observing that $\setP_{\ell}(\ell)=\{(1,\dots,\ell)\}$ consists of only a single element and after some renaming of indices, we can write \eref{e:2term4} as
\begin{equation}\label{e:2term5}
- \frac{2^{1-\ell}\rmi}{N} \sum_{a\in\mathcal{I}^{\ell}} \prs_{(1,\dots,\ell)}^{a} \sum_{b,d\in\mathcal{I}} \sum_{\scriptstyle i,j=1\atop \scriptstyle i\neq j}^\ell \eps^{a_i bd} J^{ba_j} f_{\ell-1}^{a-a_i+d-a_j}.
\end{equation}
Owing to the mutual independence of the $\prs^{a}_{p}$, we now can pick, for a given $a$, the term
\begin{equation}\label{e:term2}
- \frac{2^{1-\ell}\rmi}{N} \prs_{(1,\dots,\ell)}^{a} \sum_{b,d\in\mathcal{I}} \sum_{\scriptstyle i,j=1\atop \scriptstyle i\neq j}^\ell \eps^{a_i bd} J^{ba_j} f_{\ell-1}^{a-a_i+d-a_j}.
\end{equation}
which gives the second contribution to the time evolution equation \eref{e:schematic} for the coefficient $f_\ell^a$. Similar to the notation introduced in \eref{e:tildea}, the multi-index
\begin{equation}
a-a_i+d-a_j=(a_1,\ldots,a_{i-1},d,a_{i+1},\ldots,a_{j-1},a_{j+1},\ldots,a_\ell)
\end{equation}
in \eref{e:term2} is derived from $a$ by replacing the $i$th element by $d$ and then deleting the $j$th entry of $a$.

The contribution from the third term on the right-hand side of \eref{e:spinkin} is derived in a very similar way, so we will only sketch the calculation here. Replacing the ($\ell+1$)-spin reduced density operator $F_{1\dots\ell+1}$ by the expansion \eref{e:Fexp}, we obtain
\begin{equation}\label{e:3term_exp}
\fl \eqalign{
(N-\ell)\sum_{i=1}^\ell \Tr_{\ell+1}\pb{V_{i,\ell+1},F_{1\ldots\ell+1}}\\
\qquad= -\frac{N-\ell}{2^{\ell+1}N} \sum_{b,c\in\mathcal{I}} J^{bc} \sum_{n=1}^{\ell+1} \sum_{a\in\mathcal{I}^n} f_n^a \sum_{p\in \setP_n(\ell+1)} \sum_{i=1}^\ell \Tr_{\ell+1}\bigl[\sigma_i^b \sigma_{\ell+1}^c,\prs_p^{a}\bigr].
}
\end{equation}
Applying \eref{e:com3} to the partial trace of the commutator in \eref{e:3term_exp} one finds, similarly to the discussion of the second term, that only summands with $n=\ell+1$ can lead to terms proportional to $\prs_{(1,\ldots,\ell)}^{a}$ in \eref{e:3term_exp}. As a consequence, the sum over $p$ again consists of only a single term and, after some reshuffling of summation indices, we obtain
\begin{equation}\label{e:term3}
\frac{N-\ell}{2^{\ell-1}\rmi N} \prs_{(1,\ldots,\ell)}^{a} \sum_{b,c,d\in\mathcal{I}} J^{bc} \sum_{i=1}^\ell \eps^{a_ibd}f_{\ell+1}^{a-a_i+d+c}
\end{equation}
as the contribution to \eref{e:3term_exp} which is proportional to a given $\prs_{(1,\ldots,\ell)}^{a}$. The multi-index
\begin{equation}
a-a_i+d+c=(a_1,\ldots,a_{i-1},d,a_{i+1},\ldots,a_\ell,c)
\end{equation}
in \eref{e:term3} is obtained from $a$ by replacing the $i$th element by $d$ and then appending $c$.

This completes the proof.

\vspace{3mm}
\bibliographystyle{unsrt} 
\bibliography{paskauskas2}

\end{document}